\documentclass[twocolumn]{aastex62}
\usepackage{times}
\usepackage{epsfig,graphics}
\usepackage{amsmath, amssymb, graphicx, multirow}
\usepackage{textcomp}
\usepackage{natbib}
\usepackage{float}
\usepackage{color}
\usepackage{epstopdf}
\usepackage[countmax]{subfloat}
\usepackage[caption=false]{subfig}

\accepted{June 4, 2019}
\submitjournal{ApJ}
\shortauthors{Kalita et al.}
\begin{document}

\title{{\Large {\bf Signature of stochastic acceleration and cooling processes in an outburst phase of the TeV blazar ON 231 }}}

\correspondingauthor{Nibedita Kalita}
\email{nibeditaklt1@gmail.com}

\author{Nibedita Kalita}
\affiliation{National Astronomical Research Institute of Thailand (NARIT), Mae--Rim, Chiang Mai, 50180, Thailand}

\author{Utane Sawangwit}
\affiliation{National Astronomical Research Institute of Thailand (NARIT), Mae--Rim, Chiang Mai, 50180, Thailand}

\author{Alok C. Gupta}
\affiliation{Aryabhatta Research Institute of Observational Sciences (ARIES), Manora Peak, Nainital, 263 002, India}

\author{Paul J. Wiita}
\affiliation{Department of Physics, The College of New Jersey, P.O.\ Box 7718, Ewing, NJ 08628--0718, USA}

\begin{abstract}
We present a detailed spectral and temporal study of the intermediate-type blazar ON 231 during the TeV outburst phase in 2008 June with observations performed by {\it Swift} and {\it XMM-Newton}. The X-ray flux of the source, which was significantly dominated by the soft photons (below $3-4$ keV), varies between 27$\%$ and 38$\%$ on day time scales, while mild variations were observed in the optical/UV emissions. We found a maximum soft lag of $\sim 1$ hr between the UV and soft X-ray band which can be understood if the magnetic field of the emitting region is $\sim 5.6~ \delta^{-1/3}$ G. The $0.6-10$ keV spectra can be well represented by a broken power-law model, which indicates the presence of both synchrotron and inverse Compton components in the studied X-ray regime. The synchrotron part of the SEDs constructed with simultaneous optical/UV and X-ray data follows a log-parabolic shape. A time-resolved spectral analysis shows that the break energy varies significantly between 2.4 and 7.3 keV with the changing flux state of the source, and the similar variations of the spectral slopes of the two components support the SSC scenario. The synchrotron tail, following a log-parabolic function, shows that the peak frequency ($\nu_{p}$) varies by two orders of magnitude ($\sim 10^{14}-10^{16}$ Hz) during the event. A significantly positive $E_{p}-\beta$ relation is observed from both SED and time-resolved spectral analyses. The most feasible scenario for the observed trend during the flaring event could be associated with a magnetic-field-driven stochastic process evolving toward an equilibrium energy level.

\end{abstract}

\section{Introduction}{\label{sec:Intro}}
Blazars are the most extreme class of active galactic nuclei (AGNs) featuring relativistic jets closely aligned to the observer's line of sight \citep[$\leq$ $10^{\circ}$;][]{1995PASP..107..803U}. They are identified by strong and rapid flux variation across the entire electromagnetic spectrum, strong and variable polarization, and featureless optical spectra. Their continuum emissions are typically characterized by nonthermal radiation that is enhanced through Doppler boosting. The large amount of radiation coming out from blazars is believed to be produced by fast-moving charged particles inside an often twisted and inhomogeneous jet. 

The relativistically boosted broadband spectral energy distributions (SEDs) of blazars display two broad peaks, or humps, in the $\nu-\nu$$F_{\nu}$ representation. The low-frequency hump, which lies in the radio to soft X-ray range, is very well described by synchrotron emission from the jet, while the high-frequency hump appears in the $\gamma$-rays ranging from MeV to TeV energies \citep[e.g.][]{1995MNRAS.277.1477P}. To explain the high-frequency hump, several blazar models have been developed. Among those, the most commonly invoked scenario is the synchrotron self-Compton model \citep[SSC; e.g.,][]{1992ApJ...397L...5M, 2010ApJ...716...30A}, i.e, the inverse Compton (IC) scattering of synchrotron photons by the same electrons producing the synchrotron emission in the jet. Another widely accepted scenario is the external Compton \citep[EC; e.g.,][]{1994ApJ...421..153S} model, where external photon fields emerging from outside the jet (e.g., from the broad-line region, accretion disk, corona, or even the dusty torus or cosmic microwave background) get up-scattered by the jet electrons. These two models, SSC and EC, are collectively called leptonic models. Synchrotron radiation emitted by highly relativistic protons or muons and photopion production are another possible explanation for the high-energy hump, at least for some sources, and are known as hadronic models \citep[e.g.,][]{2013ApJ...768...54B}. Based on the peak frequency of the synchrotron hump, blazars are classified into three categories: low-synchrotron-peaked blazars (LSP), intermediate-synchrotron-peaked blazars (ISP) and high-synchrotron-peaked blazars (HSP), with those peak frequencies lying in the infrared ($\nu_{p}$ $\leq 10^{14}$ Hz), optical to near-UV ($10^{14}$ Hz $< \nu_{p} \leq 10^{15}$ Hz), and X-ray ($\nu_{p} > 10^{15}$ Hz) domains, respectively \citep[see, e.g.,][]{1995MNRAS.277.1477P, 1998MNRAS.299..433F, 2010ApJ...716...30A}. In the past couple of decades, this blazar classification has been widely used, as it seems to provide a more physical and fundamental categorization for the extragalactic objects than the alternative separation of blazars into BL Lacertae objects and flat-spectrum-radio-quasars.

With highly sensitive detectors and advanced observing facilities, a small population of blazars has been detected at extremely high $\gamma$-ray energies, $E > 100$ GeV. These blazars are popularly known as TeV blazars \citep[e.g.,][]{1998ApJ...509..608T, 2010Sci...328...73N, 2011MNRAS.414.3566T}. Initially, most of these sources were identified as high-peaked BL Lac objects at relatively low redshifts, but recently a small number of ISP and LSP blazars have been found to emit in TeV ranges \citep[e.g.,][]{2008ApJ...684L..73A, 2007ApJ...667L..21A, 2009ApJ...704L.129A, 2018MNRAS.480..879M}. Study of these particular types of blazars is important to constrain the emission mechanisms such as particle acceleration and cooling and also to indirectly probe the extragalactic background light \citep[EBL,][]{2015ApJ...812...60B}. It is only the TeV radiation that unambiguously tells us that the electrons inside the jet are accelerated to high energies (HEs) ($\sim$ TeV) with jet Doppler factors $> 10$ \citep{1995MNRAS.273..583D}. These electrons later cool off by emitting TeV photons; otherwise, $\gamma$-rays cannot escape the source owing to the severe internal photon--photon pair production process in the radiation field present near the origin \citep[e.g.,][]{2002MNRAS.336..721K, 2008MNRAS.384L..19B}. These objects are violently variable in both temporal and spectral aspects across the entire electromagnetic spectrum \citep[e.g.,][]{2009MNRAS.393L..16G, 2015ApJ...811..143P, 2016MNRAS.462.1508G}, making them prime objects of interest in exploring blazar phenomena.

ON $231$, also known as W Comae ({\it z} = $0.102$), was the first TeV-detected ISP blazar \citep{2006A&A...445..441N, 2010ApJ...716...30A}. The detection was made by {\it VERITAS} in 2008 March when the source went into a very high $\gamma$-ray outburst phase \citep{2008ApJ...684L..73A}. The broadband SED of the source in the flaring state could be described by either EC or SSC models \citep{2008ApJ...684L..73A, 2009ApJ...707..612A}. The source entered into a second outburst state in June of the same year, when the $\gamma$-ray flux was found to be higher than the previous one by a factor of three \citep{2009ApJ...707..612A}. The detection triggered a multifrequency campaign to observe the source in different wavelengths with different ground- and spaced-based telescopes. The observations that we have used in this work were carried out as part of this second multiwavelength campaign. A $237^\circ$ rotation of the optical polarization angle during the second $\gamma$-ray outburst was reported by \citet{2014ApJ...794...54S}, where the maximum estimated polarization degree was $\sim 33.3\%$. They also estimated the jet angle as $\sim 2^\circ$ and magnetic field as $\sim 0.12$ G by assuming that the photons are emitted as a result of shocks in the jet. In their study, the authors suggested that the optical- and $\gamma$-ray emitting regions could be cospatial. 


\begin{table*}
\centering
\caption{Details of {\it Swift $\&$ XMM--Newton} observations}
\small
\begin{tabular}{lllccccccc} \hline \hline 

Observatory      & Date         & ObsID     &Orbit  &Mode$^{a}$&Exposure&Pileup& Mean Counts &Flux$^{b}$ &F$_{var}$$^{c}$  \\
                 &(UT)          &            &       &       &(ks)& & (s$^{-1}$)&($\times 10^{-12}$erg cm$^{-2}$ s$^{-1}$)&(\%)\\\hline
{\it Swift}--XRT & 2008 Jun 7  & 00031219001& ...   & PC    & 8.96  & No  & 0.61$\pm$0.13 &14.62$\pm$0.43 &36.49$\pm$2.30 \\
                 & 2008 Jun 9  & 00031219002& ...   & PC    & 5.09  & No  & 0.13$\pm$0.04 &2.35$\pm$0.21  &15.73$\pm$9.09$^\star$ \\
{\it XMM--Newton}& 2008 Jun 14 & 0502211301 & 1559  &Small  & 28.20 & No  & 7.02$\pm$0.33 &12.97$\pm$0.11 &38.02$\pm$0.28 \\
EPIC-pn          & 2008 Jun 16 & 0502211401 & 1560  &Small  & 16.10 & No  & 4.67$\pm$0.28 &8.74$\pm$0.06  &27.48$\pm$0.47 \\
                 & 2008 Jun 18 & 0502211201 & 1561  &Small  & 11.40 & No  & 4.82$\pm$0.28 &9.34$\pm$0.09  &27.63$\pm$0.55 \\\hline
\end{tabular}    
Notes: $^{a}$ Operation mode $\&$ window mode used for observations, respectively, $^{b}$Absorbed 0.3--10 keV flux estimated using a power-law model, $^{c}$ Fractional variability amplitude. $^\star$ Due to the poor statistics of this observation, the estimated $F_{var}$ has a large error, and we therefore do not discuss this result in the text. 
\vspace*{0.5cm}
\end{table*}

The GHz VLBI images of the source taken in $1998$ revealed two-sided emission where bends in the jets were also observed \citep{2001A&A...374..435M}. During its optical outburst in 1998, ON 231 showed rapid flux variability on time scales of hours \citep{1999A&A...342L..49M}. In the X-ray frequencies, the source was first detected by \citet{1990ApJ...360..396W} with the {\it Einstein} satellite. After its detection, ON 231 has been observed with different X-ray satellites from time to time to investigate its continuum properties and in order to classify the object. {\it BeppoSAX} observations during that major optical outburst in 1998 revealed that the X-ray spectra have a concave shape, which was described by a broken power-law (BPL) model with upward turns at $2.5$ and $4$ keV \citep{2000A&A...354..431T}. They also observed fast flux variations below $4$ keV, but no significant variations above that energy. The authors interpreted their findings in the framework of SSC models. Similar results were reported by \cite{2005A&A...433.1163D} in 2005 with {\it BeppoSAX} observations. A recent study by \citet{2016MNRAS.458...56W} also observed the superposition of the two components in {\it Swift} X-ray telescope (XRT) spectra, where the break energy varies between $1.2$ and $4$ keV, depending weakly on the source flux state. They also found that the synchrotron flux variations are stronger than that of the Compton flux. However, in some studies, the shape of the X-ray spectra was found to follow a single power-law (PL) model without any clear upturn at HEs \citep{1997ApJ...480..534C, 2008A&A...489.1047M}. Such spectra were generally observed during fainter states of the source. In those studies, the absence of an HE upturn might be because the crossing energy of the two components lies outside the energy sensitivity limit of the satellites used for X-ray observations \citep{2000A&A...354..431T, 2008A&A...489.1047M}. 

The motivation of this work is to investigate thoroughly the X-ray spectral evolution and different variability properties of this ISP blazar in its outburst state. It is important to note that the ISP blazars are the connecting link between the high- and low-energy-peaked blazars, thus, their study might reveal information that can help us to understand the blazar phenomenon as a whole. In this paper, we present a detailed temporal and spectral study of the blazar ON 231 with {\it Swift} and {\it XMM--Newton} observations taken during the outburst phase in 2008 June. In Section 2, we briefly describe the data analysis method adopted in this work. In Sections 3 and 4, we explain the time-averaged X-ray spectral analysis and the SED analysis, respectively. Section 5 gives the temporal analysis of the event, and Section 6 includes a time-resolved X-ray spectral study. We discuss the results and summarize our findings in Section 7. 

\begin{figure}
\centering
\includegraphics[scale=0.6]{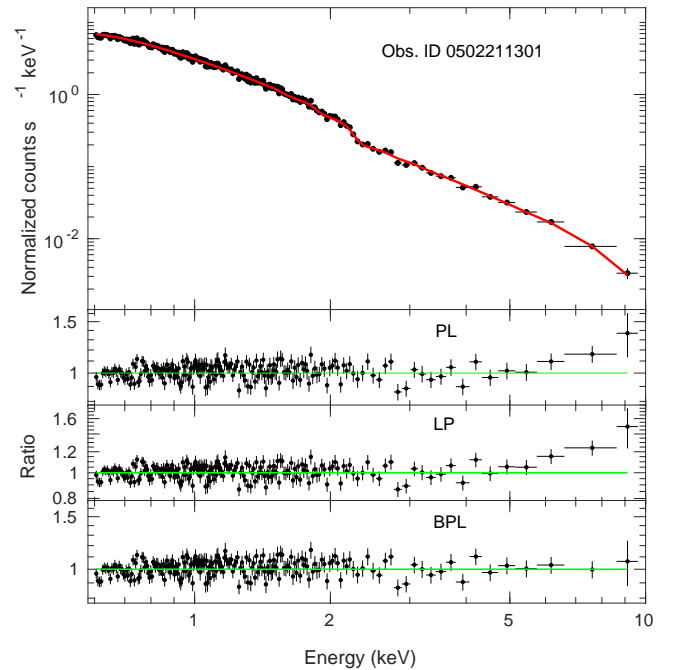}
\caption{The $0.6-10$ keV average X-ray spectra are displayed with data-to-model ratios for three different spectral models. A positive residual in the HE tail is observed while the spectra are fitted with PL or LP models. The best fit is obtained with an absorbed BPL model.}
\end{figure}

\section{Observations and Data Analysis} {\label{sec:data}}
\subsection{XMM--Newton's EPIC-pn data}

\begin{figure*}      
\centering
\mbox{\subfloat{\includegraphics[scale=0.5]{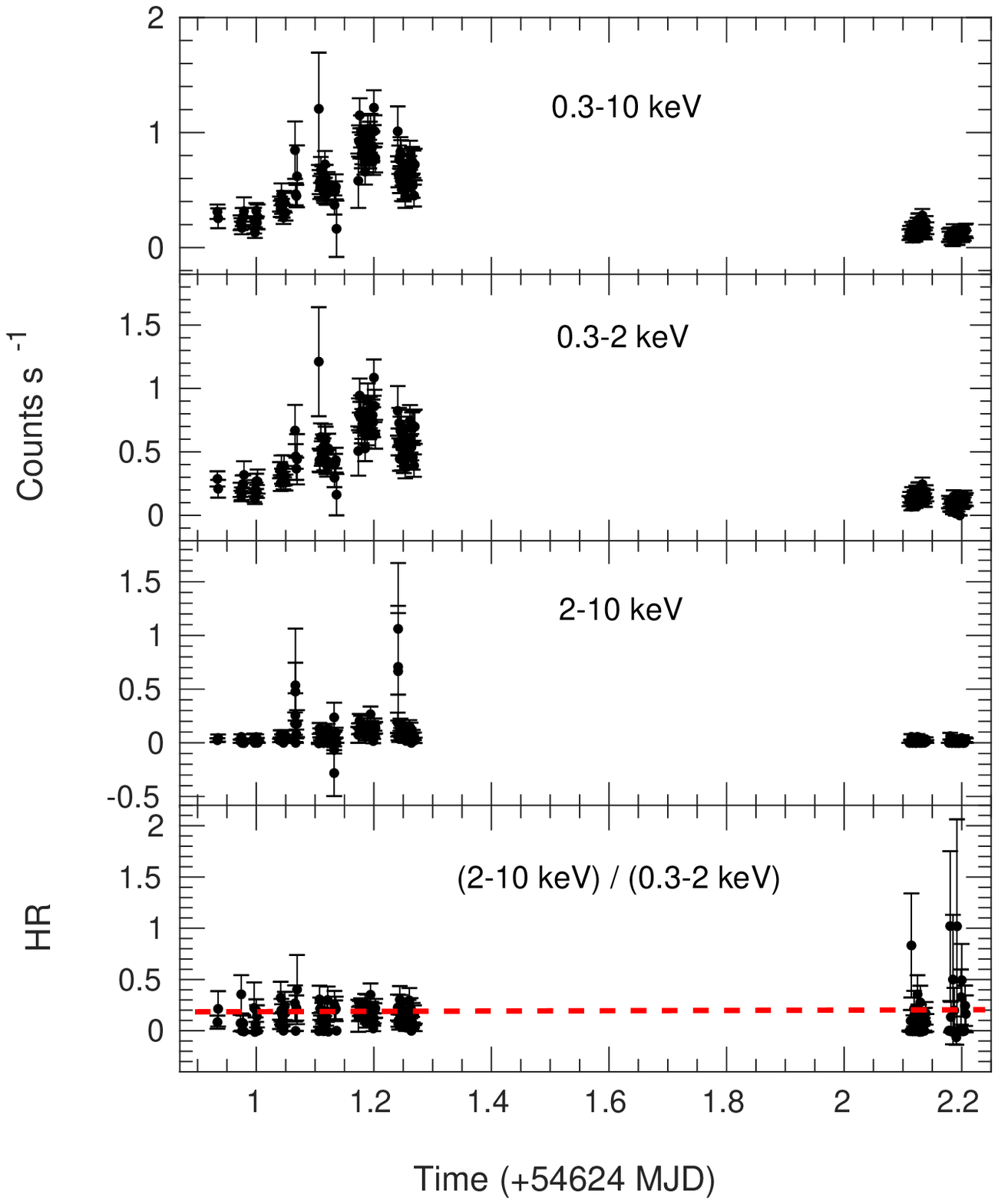}}\quad
\subfloat{\includegraphics[scale=0.5]{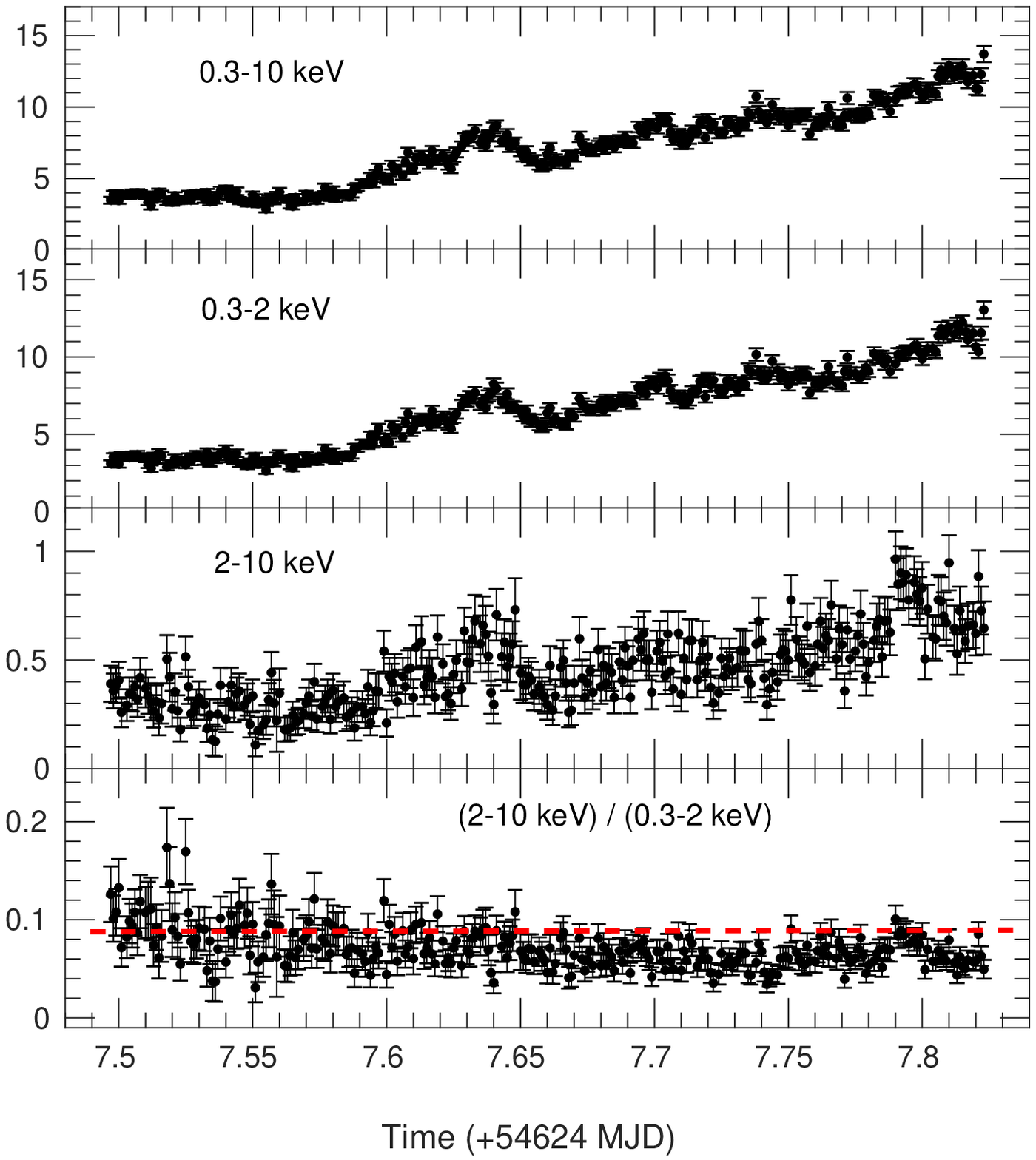} }}
\caption{Left: {\it Swift}--XRT data taken on 2008 June $7-9$. Right: {\it XMM--Newton} data taken on 2008 June 14. In both, from top to bottom, the panels show: background-subtracted total, soft-band, and hard-band LCs, and the temporal evolution of the hardness ratios, respectively, for a data bin size of $100$ s. A softer-when-brighter behavior is evident after $7.65$ ($+54624$ MJD).}
\vspace*{0.3cm}
\end{figure*}

{\it XMM-Newton} monitored the blazar on $2008$ June $14-18$ in three consecutive periods, as part of a multiwavelength campaign triggered by the detection of enhanced $\gamma$-ray emission from the source. Here we used X-ray data only from the EPIC-pn detectors owing to their better sensitivity over the MOS detectors. To reprocess the raw data files, we applied the standard pipeline by using the Science Analysis Software \citep[SAS;][]{2004ASPC..314..759G} version-17.0.0\footnote{https://www.cosmos.esa.int/web/xmm-newton/sas} and the latest calibration files\footnote{https://www.cosmos.esa.int/web/xmm-newton/calibration}. 

First, we checked the strength of high particle background on the data by extracting a light curve (LC) for the energy range of $10-12$ keV, and then we generated a data file with good time intervals free from high-background. We removed the high-background time intervals having a count rate larger than $0.4$ counts s$^{-1}$. The cleaned event files were generated by using both single and double events with the conditions $(PATTERN \leq 4)$, energy range $0.3-10$ keV, and $(FLAG $=$ 0)$ with $100$ s binning. The pileup effects were examined by using the {\it epatplot} routine, and we found that the data were unaffected by this pileup problem.

To extract the source LCs a circular region of $37^{\prime\prime}$ centered on the source was selected, and for background counts we selected a source-free circular region with a radius of $50^{\prime\prime}$ on the image, trying to avoid any contribution from the source. The final background-subtracted source counts were extracted by using the {\it epiclccorr} task. An elaborate description of the X-ray data processing we used can be found in \cite{2015MNRAS.451.1356K, 2017MNRAS.469.3824K}. In all the observations, the EPIC-pn detectors were operated using the small window mode with MEDIUM filters. The details of the observations are summarized in Table 1 and one of the resulting LCs is shown in the right panels of Figure 2. Flux corresponding to each observing epoch was measured by fitting a PL model to the $0.3-10$ keV spectra, and those values are listed in column (9) of Table 1.

For the source and background spectra, a spectral bin size of $5$ was considered with both single and double events in the energy range $0.3-10$ keV in order to have at least 25 counts per energy bin. The detector response files (rmf) were generated using the SAS task {\it rmfgen}. The final background-subtracted spectra were extracted using the task {\it grppha}.

\begin{table*}
\centering
\caption{Summary of {\it XMM--Newton's} Optical Monitor observations}
\small
\begin{tabular}{llccccccc} \hline \hline
Date      &Filter&$\lambda_{eff}^{a}$&Start$-$End Time&Exp. No.&Total Exp.&Count Rate$^{b}$&Magnitude$^{c}$&F$_{var}^{d}$ \\
(UT) &     & ($\AA$) &(hh:mm:ss)              &      & Time (ks) & (s$^{-1}$)      &              &  (\%)  \\\hline
2008 Jun 14 &U  & 3440  & 11:56:31$-$13:09:50  & 1  &4.40        &30.18$\pm$1.69& 14.549$\pm$0.004 &4.15$\pm$0.37 \\
           &UVW1 & 2910  & 13.15.02$-$19.43.11  & 5  &22.00       &14.78$\pm$1.20& 14.252$\pm$0.002 &7.68$\pm$0.22 \\    
2008 Jun 16 &U  &....   & 03.34.23$-$03.54.22  & 1  &1.20        &27.73$\pm$1.69& 14.641$\pm$0.007 &4.07$\pm$0.81 \\
           &UVW1 &....   & 03.59.36$-$08.01.12  & 5  &13.20       &12.61$\pm$1.11& 14.403$\pm$0.003 &5.29$\pm$0.38 \\
2008 Jun 18 &U  &....   & 03.26.22$-$03.46.21  & 1  &1.20        &38.58$\pm$1.98& 14.289$\pm$0.007 &3.42$\pm$0.69 \\
           &UVW1 &....   & 03.51.34$-$06.34.42  & 5  &8.50        &16.78$\pm$1.28& 14.086$\pm$0.003 &5.48$\pm$0.37\\\hline
\end{tabular}\\
Notes. $^{a}$ Effective wavelength of the filters, $^{b}$ Mean count rate, $^{c}$ Average magnitude,\\ $^{d}$ rms variability amplitude evaluated using 10 s binning of the light curves.
\end{table*}

\begin{figure*}
\centering
\includegraphics[scale=0.6]{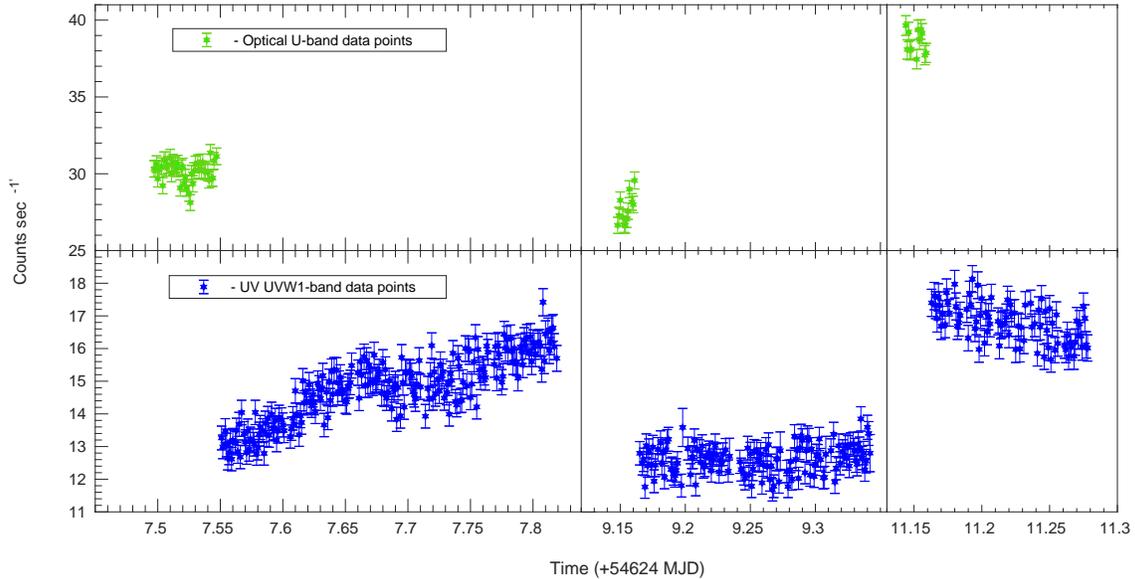}
\caption{Optical U--band and UV LCs, extracted from the fast-mode observations of XMM--Newton's OM are shown in the above plot. The left, middle, and right panels correspond to the three pointings listed in Table 2. For clear comparisons, the y-axes of optical (green) and UV (blue) LCs are kept the same for all the observations.}
\end{figure*}

\subsection{ Swift--XRT data}

We collected data from the XRT \citep[XRT;][]{2005SSRv..120..165B} on board the {\it Neil Gehrels Swift Observatory} \citep[hereafter {\it Swift};][]{2004ApJ...611.1005G}, which operates in the energy range of $0.3-10$ keV to monitor transient events. Data were reduced using the software {\it XRTDAS}, built in HEAsoft package. All the observations used here were performed in Photon Counting (PC) mode, selecting the grades $0-12$. The PC mode is generally used for faint objects having count rates $< 1$ s$^{-1}$, as was the case here. Event files were generated by applying the task {\it xrtpipeline} and using the most recent calibration files. To extract the source spectrum, we used an annulus region with an inner radius of $3$ pixels and an outer radius of $30$ pixels in order to avoid photon pileup. For the background spectrum, a source-free circular area with a radius $40$ pixels was selected, adopting these values from \citet{2009ApJ...707..612A}. Details of the {\it Swift} observations are given in Table 1 and LCs are given in the left panels of Fig.\ 2.

\subsection{Optical Monitor (OM) data}

The OM \citep{2001A&A...365L..36M} on board {\it XMM--Newton} was operated in both imaging and fast photometry modes during the whole event. The imaging mode is especially used for obtaining the spatial resolution of an X-ray object. On the other hand, the fast mode is used for collecting optical/UV photons with high time resolution, and so it is very useful for monitoring rapidly variable sources such as blazars.

Several exposures were collected in the optical band, U ($\lambda_{range} = 3000-3900$ \AA) and in the UV band ($\lambda_{range} = 2450-3200$ \AA). We reprocessed these data with the perl script {\it omichain} to get the observation source list that contains the calibrated photon counts and their corresponding errors. We extracted the fast-mode time-series data by using the task {\it omfchain}. The default binning was 10 s, which was used to estimate the strength of the variability. For the optical and UV LCs plotted in Fig.\ 3, the binning was changed to 100 s for comparison to the X-ray LCs binned at that interval. The optical/UV spectra were extracted from photometric observations using the task {\it om2pha} in order to analyze in {\it XSPEC}. The exact coordinates of the source were determined by performing interactive photometry in OM images using the tool {\it omsource}. These spectra were used in broadband SED analysis with the corresponding canned response files for each filter. The details of the optical/UV exposures are summarized in Table 2 and the LCs are shown in Figure 3.

\begin{table*}
\centering
\caption{Result of time-averaged spectral analysis (galactic absorption is fixed at N$_{H}$ = 2.18 $\times$ 10$^{20}$ cm$^{-2}$)}
\small
\begin{tabular}{llclcccc} \hline \hline
Observing& Model  &$\Gamma/\alpha/\Gamma_{1}^{a}$ & E$_{break}^{b}$ &$\beta/\Gamma_{2}^{c}$   & N$^{d}$     &$\chi_{r}^{2}$(dof) &  $F_{0.6-10}^{e}$ \\
 Date    &        &                  &   (keV)       &                &($\times$ 10$^{-3}$)   && ($\times$ 10$^{-12}$ erg cm$^{-2}$ s$^{-1}$)\\\hline

07/06/2008& PL    & 2.65$\pm$0.07  &...                    & ...                    & 3.64$\pm$0.13 &0.74 (388) &  10.06$\pm$0.34 \\
          & LP    & 2.55$\pm$0.12  &...                    & 0.27$_{-0.26}^{+0.28}$ & 3.69$\pm$0.14 &0.73 (387) &  9.77$\pm$0.45\\
          & BPL   & 2.66$\pm$0.07  &4.26$_{-0.64}^{+0.73}$ & 2.07$_{-0.73}^{+0.82}$ & 3.64$\pm$0.21 &0.74 (387) &  10.38$\pm$0.33 \\\hline  

09/06/2008& PL    & 2.80$\pm$0.28  &...                    & ...                    & 0.55$\pm$0.49 &0.73 (162) &  1.42$\pm$0.18 \\
          & LP    & 2.95$\pm$0.40  &...                    &-0.44$_{-0.78}^{+0.96}$ & 0.54$\pm$0.04 &0.73 (161) &  1.51$\pm$0.39\\
          & BPL   & 2.89$\pm$0.33  &2.38$_{-1.20}^{+1.45}$ & 2.37$_{-0.81}^{+1.05}$ & 0.55$\pm$0.21 &0.74 (160) &  1.49$\pm$0.31 \\\hline

14/06/2008& PL    & 2.75$\pm$0.01  &...                    & ...                    & 3.13$\pm$0.02 &1.04 (542) &  8.23$\pm$0.07 \\
          & LP    & 2.73$\pm$0.02  &...                    & 0.06$_{-0.05}^{+0.05}$ & 3.15$\pm$0.02 &1.03 (541) &  8.18$\pm$0.09\\
          & BPL   & 2.76$\pm$0.01  &5.33$_{-1.07}^{+2.25}$ & 2.26$_{-1.78}^{+0.33}$ & 3.13$\pm$0.01 &1.03 (540) &  8.32$\pm$0.13 \\\hline
16/06/2008& PL    & 2.86$\pm$0.03  &...                    & ...                    & 2.06$\pm$0.02 &1.21 (370) &  5.16$\pm$0.08\\ 
          & LP    & 2.89$\pm$0.03  &...                    &-0.15$_{-0.09}^{+0.09}$ & 2.02$\pm$0.03 &1.20 (369) &  5.26$\pm$0.10\\ 
          & BPL   & 2.89$\pm$0.03  &2.68$_{-0.32}^{+0.45}$ & 2.48$_{-0.17}^{+0.14}$ & 2.05$\pm$0.01 &1.17 (368) &  5.35$\pm$0.11\\\hline

18/06/2008& PL    & 2.75$\pm$0.03  &...                    & ...                    & 2.22$\pm$0.01 &1.02 (344) &  5.82$\pm$0.08\\ 
          & LP    & 2.84$\pm$0.04  &...                    &-0.31$_{-0.09}^{+0.09}$ & 2.17$\pm$0.03 &0.95 (343) &  6.09$\pm$0.13\\     
          & BPL   & 2.79$\pm$0.03  &3.87$_{-0.51}^{+0.67}$ & 1.73$_{-0.56}^{+0.35}$ & 2.21$\pm$0.04 &0.93 (342) &  6.22$\pm$0.15\\\hline 
\end{tabular}     \\
Note: $^{a}$ Spectral index for PL/LP/BPL, $^{b}$ Break energy between the two components, $^{c}$ Curvature parameter of \\ LP/ second SI of BPL fit,
$^{d}$ PL normalization, $^{e}$ Spectral flux in the energy range of 0.6--10 keV.\\
\vspace*{0.3cm}
\end{table*}

\section{Time-averaged Spectra}

We performed the X-ray spectral analysis using {\it XSPEC} version 12.8.2 software \citep{1996ASPC..101...17A}, available with the package HEAsoft. We used the $\chi^{2}$ statistic to determine the goodness of fits to spectra. We performed the spectral analysis over the energy range $0.6-10$ keV owing to the higher quantum efficiency of the EPIC-pn detector in this range \citep{2001A&A...365L..18S}. Unless stated otherwise, all the errors quoted are $2\sigma$ ($90\%$).

In order to fit the X-ray spectra, we first used a simple PL model with Galactic absorption. The Galactic hydrogen column density was fixed at the neutral {\it H} value, $N_{H}$ $= 2.18 \times 10^{20}$ cm$^{-2}$ \citep{2013MNRAS.431..394W}. Although we got acceptable fit statistics, a positive data-to-model ratio was present above $3-4$ keV in all the observations, indicating that a PL model was not sufficient to describe the spectra. The positive HE tails in the residual plots are clear indications of the spectra being intrinsically curved. In the case of blazars, an upward bending of X-ray spectra is most simply interpreted in terms of synchrotron and IC components both being present and making comparable contributions. Such spectra allow us to investigate properties of both the emitting electron distributions \citep{2000A&A...354..431T, 2005A&A...433.1163D}.
 
To fit the curved spectra, we adopted two models: a log-parabolic (LP) PL and a BPL model. In both cases, we fixed the $N_{H}$ value as in the PL fitting. An LP model could not remove the positive bending of the HE tail. However, the BPL model significantly improved the fitting by giving equally distributed residual points around the fixed reference line. The spectral fits with residuals corresponding to the three models are shown in Figure 1 and Figure 13, and the fit statistics are listed in Table 3. We found that the break energy ($E_{break}$) evaluated from BPL fitting covers a wide range of energy, from $2.38$ to $5.33$ keV. We also found that the synchrotron spectra generally are steeper than the IC spectra, where their two spectral slopes, $\Gamma_{1}$ and $\Gamma_{2}$, varied between $2.66\pm 0.07$ to $2.89 \pm 0.03$ and $1.73^{+0.35}_{-0.56}-2.48^{+0.19}_{-0.17}$, respectively.

From the above analysis, we observed that the PL and LP models can fit the spectra below the break, but it was not clear which model better represents the synchrotron spectra. So although the BPL model defined the averaged spectra, the particles producing the synchrotron spectrum might follow a curved function. To investigate this possibility, we extended the {\it XMM-Newton} X-ray spectra up to the optical wavelength using OM data and performed spectral fitting, which is discussed in the next section.

\section{ SED}

We constructed the SEDs at optical/UV to X-ray frequencies using simultaneous EPIC-pn and OM observations. In order to investigate the true nature of the electron distribution, we fitted the synchrotron part of the SED from optical (U-filter, OM) up to the break energy with a simple PL model and an LP model. We found that the spectra are intrinsically curved instead of having a straight slope and were well fitted by an LP function. The LP model \citep{2004A&A...413..489M} which represents an energy-dependent curved spectrum, is usually described as

\begin{equation}
F(E) = N \Bigl(\frac {E}{E_{0}} \Bigr)^{-[ \alpha+\beta {\rm log}(E/E_{0})]} . 
\end{equation}

The reference energy, $E_{0}$, at which the model normalization, $N$, and photon index, $\alpha$, are estimated is fixed at 1 keV. The measure of model curvature, $\beta$, can be used to estimate other quantities of the physical model such as the energy-dependent photon index and the synchrotron peak frequency, $\nu_{p}$, corresponding to the maximum flux, $\nu_{p}F(\nu_{p})$ (in  units of erg cm$^{-2}$ s$^{-1}$), of the SED as follows:

\begin{figure}
\hspace*{-1.0cm}
\centering
\includegraphics[scale=0.28, angle=-90]{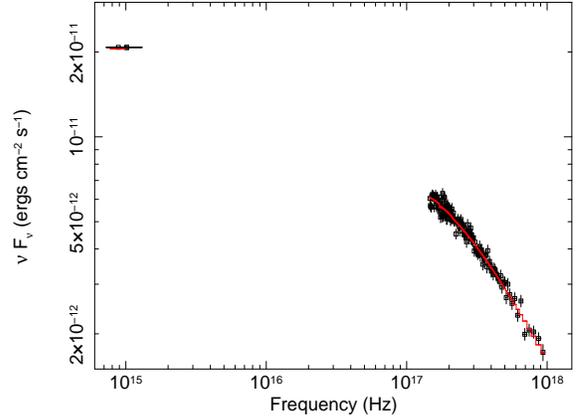}
\caption{The SED of ON 231 in outburst phase with simultaneous optical/UV and X-ray observations on 2008 June 14. The average SED is unfolded with an LP model, where the upper limit for the X-ray frequency comes from the break energy, $E_{break}$, of the average spectra.}
\end{figure}

\begin{figure}
\centering
\includegraphics[scale=0.5]{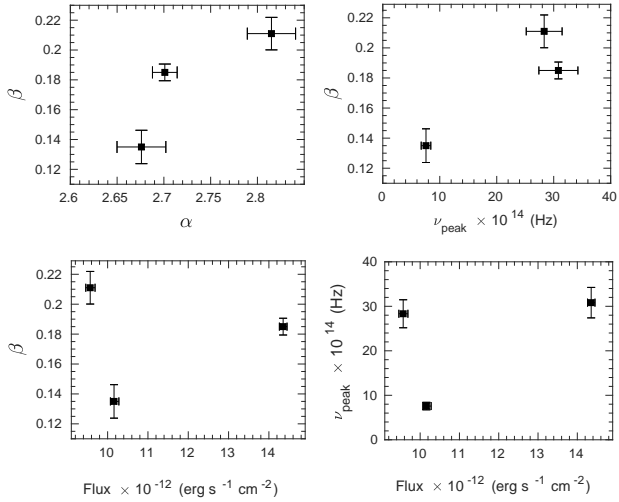}
\caption{Relations between SED parameters for the three {\it XMM--Newton} observations. Top panels: curvature parameter against spectral slope (left) and peak frequency (right). Bottom panels: spectral curvature (left) and peak frequency (right) as a function of model flux.}
\end{figure}

\begin{table*}
\centering
\caption{Summary of broadband SED fits using a log-parabolic model}
\begin{tabular}{lcccccrcc} \hline \hline
ObsID  &$\alpha$ &$\beta$ & N &  $\chi^2_r$ (dof)  & $F_{0.05-10}^{a}$  & $E_{peak}^{b}$  & $\nu_{p}$ $F^{c}(\nu_{p})$ &$F_{bol}^{d}$ \\
      &         &        & ($\times 10^{-3}$)&                & ($\times 10^{-12}$erg cm$^{-2}$ s$^{-1}$) &  (eV)    &(erg cm$^{-2}$ s$^{-1}$)&(erg cm$^{-2}$ s$^{-1}$)  \\\hline

0502211301 &2.70$\pm$0.01 &0.19$\pm$0.01 &3.16 &1.04 (462)  &14.35$\pm$0.09 &12.75 &2.34$\times 10^{-11}$ &1.47$\times 10^{-10}$\\ 
                                                                                              
0502211401 &2.82$\pm$0.03 &0.21$\pm$0.01 &2.07 &1.48 (304)  &9.57 $\pm$0.11 &11.72 &2.03$\times 10^{-11}$ &1.19$\times 10^{-10}$\\
                                                                                              
0502211201 &2.68$\pm$0.03 &0.14$\pm$0.01 &2.21 &1.44 (308)  &10.16$\pm$0.12 &3.14  &2.48$\times 10^{-11}$ &1.82$\times 10^{-10}$\\\hline
\end{tabular} \\                                                       
Notes:  $^{a}$ SED flux in the energy range 0.05--10 keV estimated from LP fitting,
 $^{b}$ Peak energy corresponding to the maximum \\ flux, $\nu_{p}$ $F(\nu_{p})$, of the SED,
 $^{c}$ Maximum flux of the SED,
 $^{d}$ Bolometric flux corresponding to $\nu_{peak}$. \\
\vspace*{0.3cm}
\end{table*}

\begin{equation}
E_{p} = E_{0} 10^{(2-\alpha)/2\beta}, 
\end{equation} 
and 
\begin{equation}
\nu_{p}F(\nu_{p}) = (1.60 \times 10^{-9}) NE_{0}E_{p}\Bigl(\frac{E_{p}}{E_{0}}\Bigr)^{-\alpha/2}.
\end{equation} 

Results of the SED analysis are summarized in Table 4 and an example of the SED fitting is shown in Figure 4. The peak energies, $E_{p}$ of the SEDs were estimated in different epochs using the Eq.\ (2). Several SED parameters are plotted against each other in Figure 5 and are discussed in Section 7.

\section{Temporal Analysis}
X-ray LCs in the energy range 0.3--10 keV, extracted from {\it Swift}--XRT and {\it XMM--Newton}'s EPIC-pn pointings, are plotted in Fig.\ 2. The soft-band (0.3--2 keV) and hard-band (2--10 keV) LCs along with the hardness ratio (HR), evaluated as HR = hard band count rates / soft band count rates, are also plotted in the same figure. The optical and UV LCs extracted from OM are shown in Figure 3, in the top and bottom panels, respectively. The vertical solid lines dividing the plot denote the gaps between the observations (see Table 2).

\subsection{Flux Variability}
The observations listed in Tables 1 and 2 were used to study the temporal variability of the flaring event. The strength of variability was estimated using the fractional rms variability amplitude ($F_{var}$) method given by \citet{2002ApJ...568..610E}. The average variability amplitude with respect to the mean flux of an object is given by 

\begin{figure}
\centering
\includegraphics[scale=0.47]{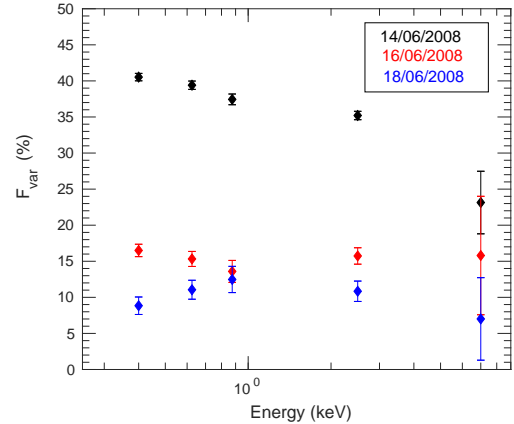}
\caption{Fractional variability amplitude, $F_{var}$, as a function of photon energy for the {\it XMM--Newton} observations. Different patterns of variability are seen during different epochs. }
\end{figure}

\begin{equation}
F_{var} = \sqrt{ \frac{ S^{2} -
\overline{\sigma_{err}^{2}}}{\bar{x}^{2}}}
\end{equation}
where $\overline{\sigma_{err}^{2}}$ is the mean square error and $S^2$ is the sample variance of the time-series data. The uncertainty on $F_{var}$ has been calculated using the Monte Carlo method for a total $n$ number of photon measurements as described in \citet{2003MNRAS.345.1271V}. 

The values estimated using the above equation are tabulated in the last columns of Tables 1 and 2. The $F_{var}$ during the event varied between 27$\%$ and 38$\%$ with a 100$\%$ duty cycle for the variations.

\begin{figure*} 
\centering
\mbox{\subfloat{\includegraphics[scale=0.55]{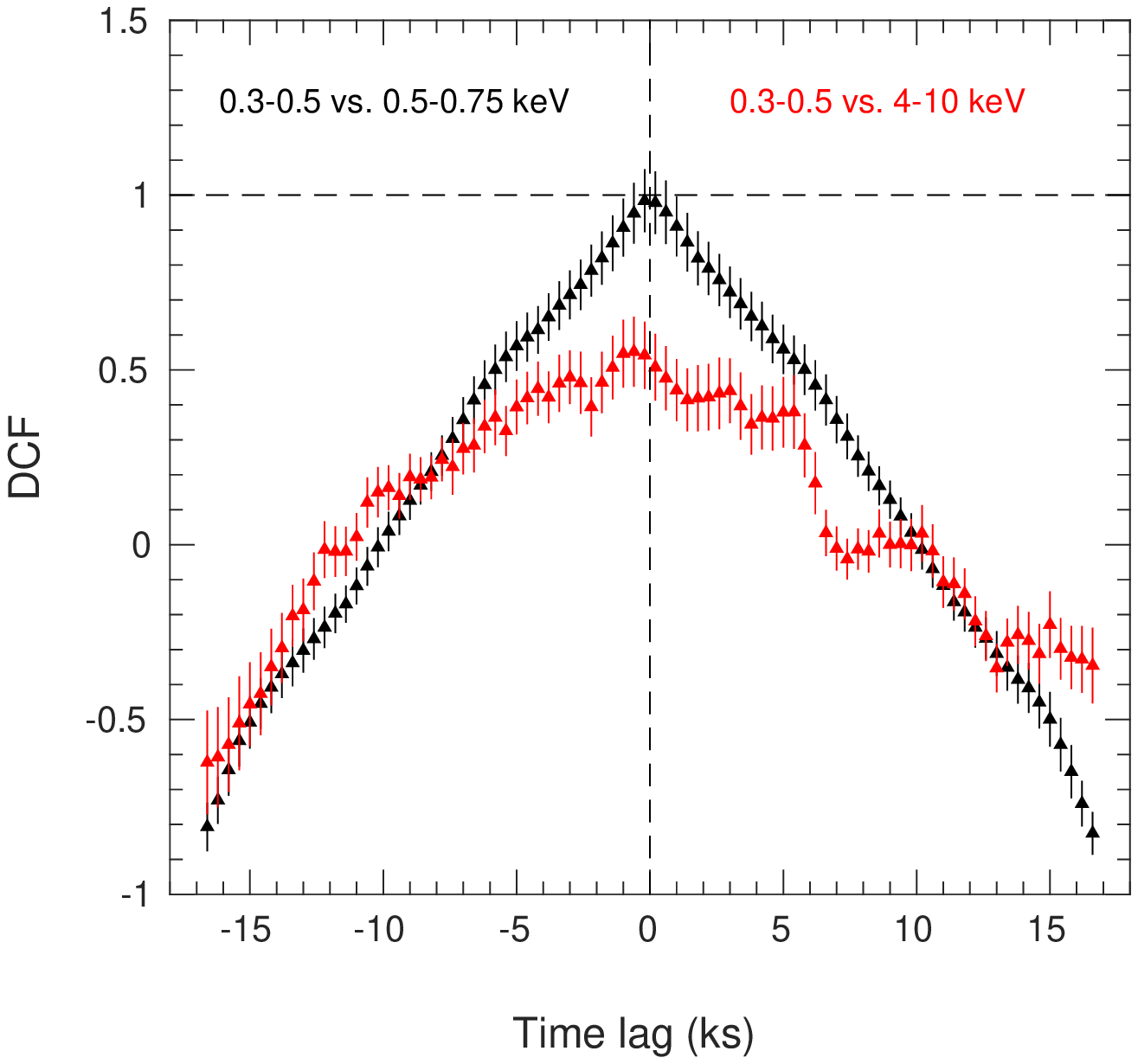}}\quad
\vspace*{2.5cm}
\subfloat{\includegraphics[scale=0.55]{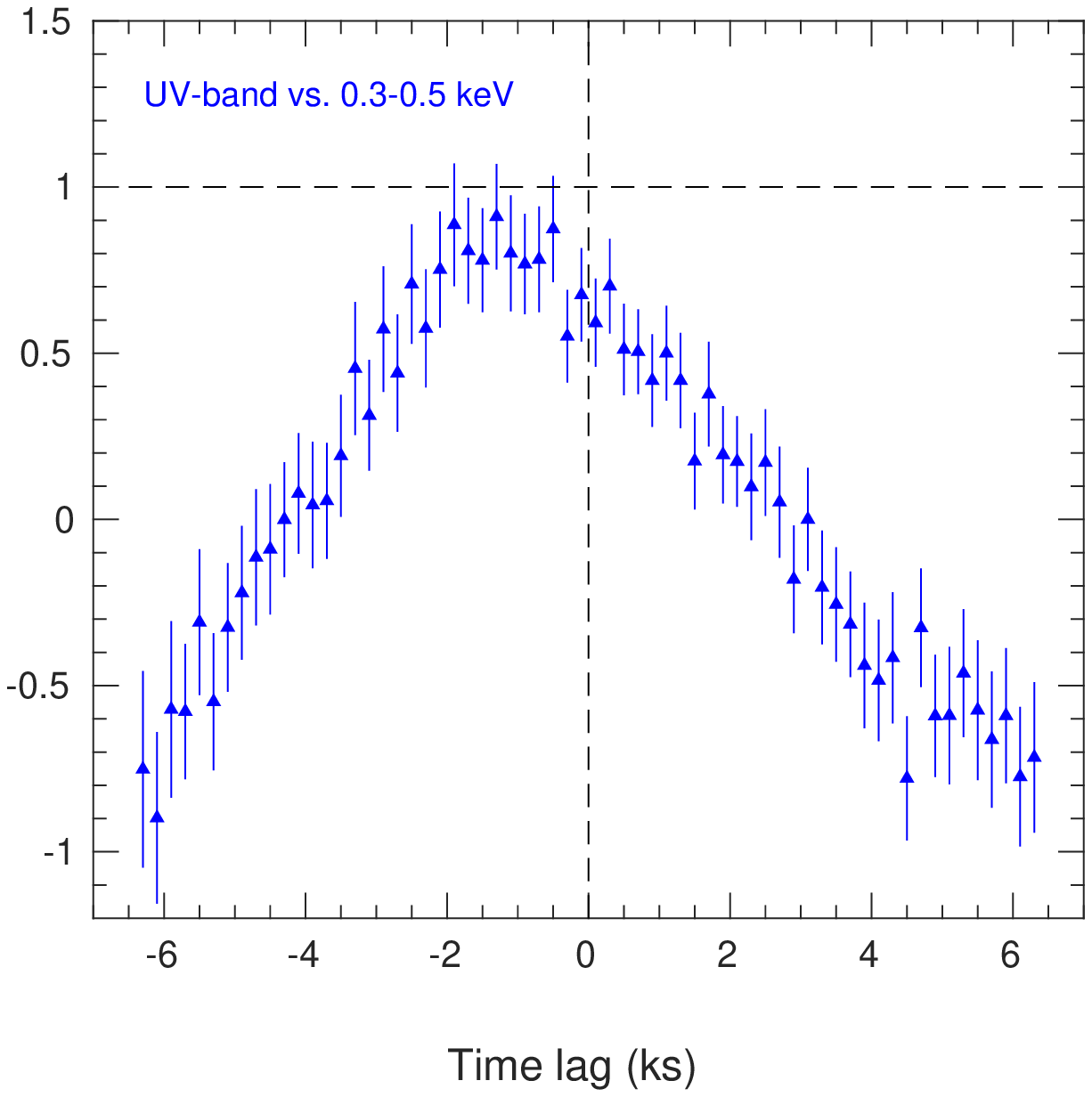} }}
\caption{Left: DCF between different X-ray energy bands. A soft lag of $\sim -0.40$ ks is detected between the 0.3--0.5 keV and 4--10 keV bands (red curve). Right: DCF representation between the UV and 0.3--0.5 keV X-ray bands, where a soft lag around $-1.25$ ks is observed.}
\end{figure*}

\subsection{Discrete Correlation Function (DCF) Analysis}

Based on the break energy derived from time-averaged spectral fits, we first split the {\it XMM--Newton} X-ray LCs into two main energy bands: 0.3--4 keV and 4--10 keV, which are respectively dominated by the synchrotron and IC components. Later, in order to study correlated variability, the low-energy band was further divided into four bands: 0.3$-$0.5, 0.5$-$0.75, 0.75$-$1, and then 1$-$4 keV \citep[following][]{2010ApJ...713..180Z}. During the whole event, the overall X-ray emission was significantly dominated by the soft X-ray photons, with mean counts $\sim 10$ times greater than that of hard emission (IC component), and this ratio decreased as the outburst decayed. During the whole event, the synchrotron flux varied between $\approx$ 41$\%$ and 9$\%$ while the IC flux varied between 23$\%$ and 7$\%$. We notice that the difference of variability between the soft (synchrotron) and hard (IC) emissions also decreases with time during the decay of the X-ray flare, becoming comparable toward the end of the event. The flux variability, $F_{var}$ as a function of photon energy is plotted in Figure 6 which shows different patterns for all three {\it XMM--Newton} observations. 


\begin{table}
\caption{Summary of DCF analysis}
\small
\centering
\begin{tabular}{lllcc} \hline \hline
ObsID & Energy Band (keV) &DCF$_{peak}$  &  Time Lag (ks)   \\\hline

0502211301&UVW1       & 0.87   & -3.25  $_{-1.12 }^{+ 1.09}$ \\
       &4$-$10        & 0.55   & -0.91  $_{-0.37 }^{+ 0.34}$ \\
                                                                                      
0502211401&0.5$-$0.75      & 0.99   &  0.12  $_{-0.05 }^{+ 0.06}$ \\
     &0.75$-$1        & 0.95   &  0.13  $_{-0.06 }^{+ 0.06}$ \\
     &1$-$4           & 0.87   &  0.39  $_{-0.09 }^{+ 0.09}$ \\
                                                                                    
0502211201&UVW1       & 0.88   & -1.25 $_{-0.51  }^{+ 0.37}$\\
     &0.5$-$0.75      & 0.96   & -0.50 $_{-0.20  }^{+ 0.15}$\\
     &0.75$-$1        & 0.95   & -0.72 $_{-0.33  }^{+ 0.35}$\\
     &1$-$4           & 0.90   & -0.75 $_{-0.21  }^{+ 0.15}$\\\hline
\end{tabular}   \\                                                     

Note. All the DCF values are estimated with respect to the $0.3-0.5$ keV band. DCF$_{peak}$ and time lags are estimated by fitting a Gaussian curve to the DCF data points. In the above\\ table, we have shown the result for only those measurements\\ for which DCF$_{peak}$ $>$ 0.5 or $\tau$$>$ bin size (100 s).
\end{table}

The Discrete Correlation Function \citep[DCF;][]{1988ApJ...333..646E} is one of the best methods to investigate a correlation between two unevenly sampled time-series data. An advantage of using this method is that it is capable of handling unevenly sampled data without interpolating it, thus giving more accurate results with proper error estimation. If we consider two discrete data trains, $x_{i}$ and $y_{i}$, then for each data pair ($x_{i}$, $y_{i}$), with $0 \leq i,j \leq n$, the unbinned DCF (UDCF) is given by

\begin{equation} 
UDCF_{ij}(\tau) = \frac{(x_i-\bar{x})(y_j-\bar{y})}{\sqrt{\sigma_{x}{^2} \sigma_{y}{^2}  }}    ,
\end{equation}
where $n$ is the number of data points and the other symbols have their usual meanings.
Here, each value of $UDCF_{ij}(\tau)$ is associated with the time delay, \(\Delta t_{ij} = (t_{j}-t_{i}) \). The final DCF values are measured by averaging the UDCF values over $m$ number of pairs that lie in the range \( \tau - \frac{\Delta\tau}{2} \leq t_{ij} \leq \tau+ \frac{\Delta\tau}{2} \) and are computed through

\begin{equation}
 DCF{(\tau)}= \frac{1}{m}{\sum_{k=1}^m UDCF_{k}}.
\end{equation}

A sensible measurement of DCF depends on correct estimation of the mean and variance of the time-series data. In general, these parameters are estimated from the entire data set, which will be valid only if the data sets used in the calculation are statistically stationary. However, this is most unlikely for AGN LCs \citep{1994PASP..106..879W}. Thus, for safer estimation of mean and variance, we used only the data points that contribute to the calculation of the DCF at any particular $\tau$. If the maximal value of the DCF, DCF$_{peak}$, is $> 0$, then the two data sets are positively correlated, and a higher peak value represents a higher degree of correlation; however, DCF$_{peak} <$ 0 represents anticorrelations and DCF$_{peak}$ $= 0$ indicates the absolute absence of any correlation between the data sets. 

\begin{figure*}
\centering
\includegraphics[scale=0.65]{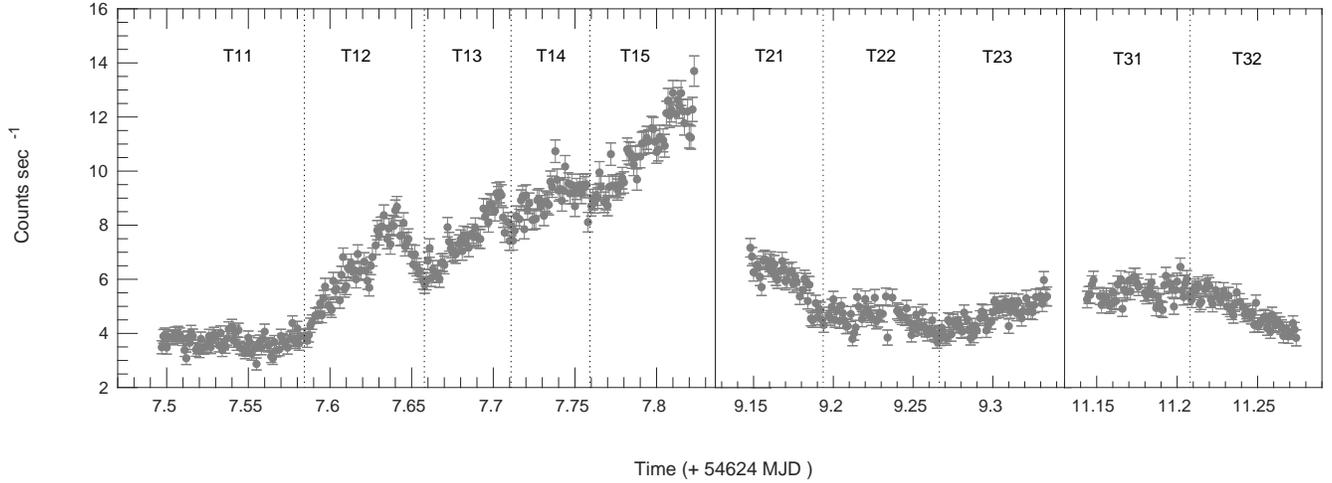}
\caption{Temporal evolution of the X-ray outburst event of ON 231 recorded by {\it XMM--Newton} in $2008$ June. The total LC combines three pointings separated by the solid lines, while the dotted lines delineate the individual episodes used for the time-resolved spectral analysis.} 
\end{figure*}

We used the above method for searching for possible correlations between the optical/UV and X-ray bands and also between the X-ray sub-bands. To do so, we measured DCFs between the $0.3-0.5$ keV LC and all the other energy bands. The DCF for $0.3-0.5$ keV vs.\ $0.5-0.75$ keV showed a peak at $\approx$ 1 with zero time lag in all the observations, as shown in Figure 7 (black curve). The DCFs for $0.75-1$ keV and $1-4$ keV follow the same pattern. There is a weak correlation between the $0.3-0.5$ and $4-10$ keV bands, as also shown in Fig.\ 7 (red curve), but with a modest but detectable temporal lag. The results of our DCF analyses are summarized in Table 5. The peak values and time lags were estimated by fitting a Gaussian model to the DCF measurements. The original LCs are binned at $100$ s and the DCF measurements are made at $300$ s intervals, so any time lags having values less than $100$ s are not considered as significant detection. 

From this analysis, we found both soft and hard lags during the flaring event. The maximum time delay was detected between the UV and soft X-ray bands, having a negative lag of $\approx 1$ hr. In Figure 7, the DCF plots for the soft X-ray band vs. UV, soft- and hard-energy bands are shown. For clearer visualization of the DCF peak and any corresponding lag, vertical and horizontal dashed lines are drawn at $\tau = 0$ and DCF = $1$, respectively.

\section{Time-resolved Spectra}
Integrated spectra over a long exposure or covering different and changing flux levels might give misleading results. Thus, in order to examine the evolution of the X-ray flares, we performed a time-resolved spectral analysis with the {\it XMM--Newton} observations. 

The combined time-series data from {\it XMM--Newton} are shown in Figure 8, from which it is clearly visible that there were different flux states during the burst phase of the source. In the figure, the solid lines represent temporal gaps between the EPIC-pn observations. For comparison, the flux-axis limits are kept the same for all the observations. The first observation covers the steep rising phase of the main outburst event, while the second observation records the decaying part of the event and rising part of a mild (moderate) flare, and the third pointing starts at a nearly constant level before decaying. All of the time-series data were sliced into intervals based on their changing flux patterns, which gave us 10 epochs, shown separated by vertical dotted lines in Fig.\ 8. Each interval represents a different phase of the event; it can represent a stable flux state as seen in the intervals T11 $\&$ T31, or a mini X-ray flare (T12, T13, T14 $\&$ T22), or a rising (T15 $\&$ T23) or decaying (T21 $\&$ T32) part of high- or low-amplitude flares.


\begin{table*}
\centering
\caption{Results of time-resolved spectral fitting of the $0.6-10$ keV spectra with a broken power-law model}
\small
\begin{tabular}{llcccccccccc} \hline \hline
ObsID&Interval &$\Gamma_{1}^{a}$ &$E_{break}^{b}$ &$\Gamma_{2}^{c}$ &$N^{d}$  &$\chi_{r}^{2}$(dof)&$F_{total}^{e}$& $F_{sync1}^{f}$   &  $F_{sync2}^{g}$ \\
     &     &                 &  (keV)         &        &($\times$ 10$^{-3}$)&&($\times$ 10$^{-12}$)&($\times$ 10$^{-12}$)&($\times$ 10$^{-12}$)\\\hline
0502211301&T11& 2.61$_{-0.04}^{+0.04}$&4.83$_{-0.89}^{+1.32}$& 1.63$_{-0.91}^{+0.56}$&1.64 &1.02 (256) &4.88 $\pm$0.17& 3.85$\pm$0.07 &4.64$\pm$0.13\\
          &T12$^\star$& 2.44$_{-0.17}^{+0.12}$&0.89$_{-0.11}^{+0.19}$& 2.77$_{-0.09}^{+0.08}$&3.03 &0.92 (289) &7.48 $\pm$0.46& 6.44$\pm$0.09 &7.58$\pm$0.12\\
          &T13& 2.82$_{-0.03}^{+0.03}$&4.52$_{-0.54}^{+0.60}$& 2.38$_{-0.43}^{+0.51}$&3.29 &0.92 (254) &8.54 $\pm$0.55& 7.31$\pm$0.13 &8.41$\pm$0.19\\
          &T14& 2.83$_{-0.03}^{+0.03}$&5.30$_{-1.31}^{+1.21}$& 2.33$_{-1.01}^{+1.06}$&3.89 &1.05 (282) &10.01$\pm$0.12& 8.63$\pm$0.12 &9.73$\pm$0.17\\
          &T15& 2.79$_{-0.03}^{+0.03}$&7.31$_{-1.21}^{-1.21}$& 0.94$_{-0.68}^{+0.41}$&4.87 &0.96 (306) &12.68$\pm$0.06&10.86$\pm$0.11&12.53$\pm$0.12\\

0502211401&T21& 2.93$_{-0.05}^{+0.06}$&2.41$_{-0.84}^{+0.84}$& 2.83$_{-1.01}^{+1.04}$&2.41 &0.93 (190) &5.96 $\pm$0.09& 5.24$\pm$0.11 &5.91$\pm$0.21\\
          &T22& 2.95$_{-0.05}^{+0.05}$&2.80$_{-0.44}^{+0.92}$& 2.31$_{-0.47}^{+0.31}$&1.85 &1.00 (216) &4.84 $\pm$0.19& 4.01$\pm$0.12 &4.51$\pm$0.13\\
          &T23& 2.82$_{-0.05}^{+0.03}$&2.63$_{-0.54}^{+0.76}$& 2.35$_{-0.26}^{+0.19}$&1.93 &1.08 (228) &5.30 $\pm$0.22& 4.31$\pm$0.08 &4.95$\pm$0.13\\

0502211201&T31& 2.79$_{-0.04}^{+0.04}$&3.85$_{-1.09}^{+1.01}$& 1.82$_{-0.77}^{+0.50}$&2.35 &1.04 (275) &6.56 $\pm$0.27& 5.26$\pm$0.09 &6.09$\pm$0.11\\
          &T32& 2.78$_{-0.05}^{+0.06}$&3.80$_{-0.89}^{+1.15}$& 1.74$_{-0.89}^{+0.49}$&1.98 &0.77 (188) &5.58 $\pm$0.16& 4.48$\pm$0.12 &5.21$\pm$0.14\\\hline
\end{tabular}  \\  
  Notes. 
  $^{a}$Photon index of first PL. 
  $^{b}$Break energy between the two emission components. 
  $^{c}$Photon index of second PL. \\
  $^{d}$Model normalization in units of photons keV$^{-1}$ cm$^{-2}$ s$^{-1}$ estimated at 1 keV.
  $^{e}$0.6--10 keV model flux in units of erg cm$^{-2}$ s$^{-1}$. \\
  $^{f}$0.6--4 keV Synchrotron flux in units of erg cm$^{-2}$ s$^{-1}$.
  $^{g}$Extended 0.6--10 keV synchrotron flux in units of erg cm$^{-2}$ s$^{-1}$.\\
  $^\star$ $E_{b}$ could not be well constrained for interval T12, so we do not quote any result from this fit in the text.
\end{table*}

\begin{figure*} 
\centering

\mbox{\subfloat{\includegraphics[scale=0.4]{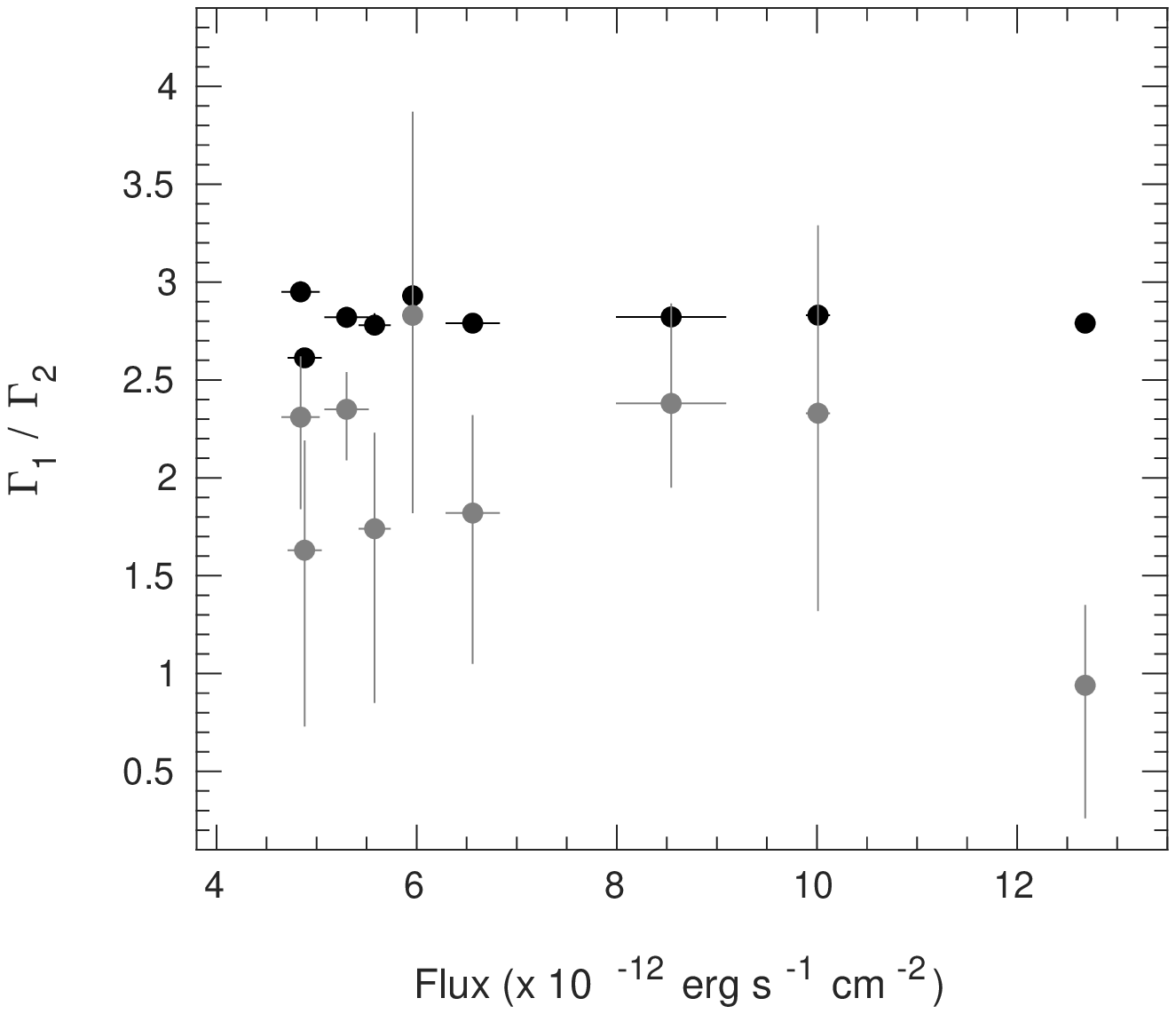}}\quad

\subfloat{\includegraphics[scale=0.4]{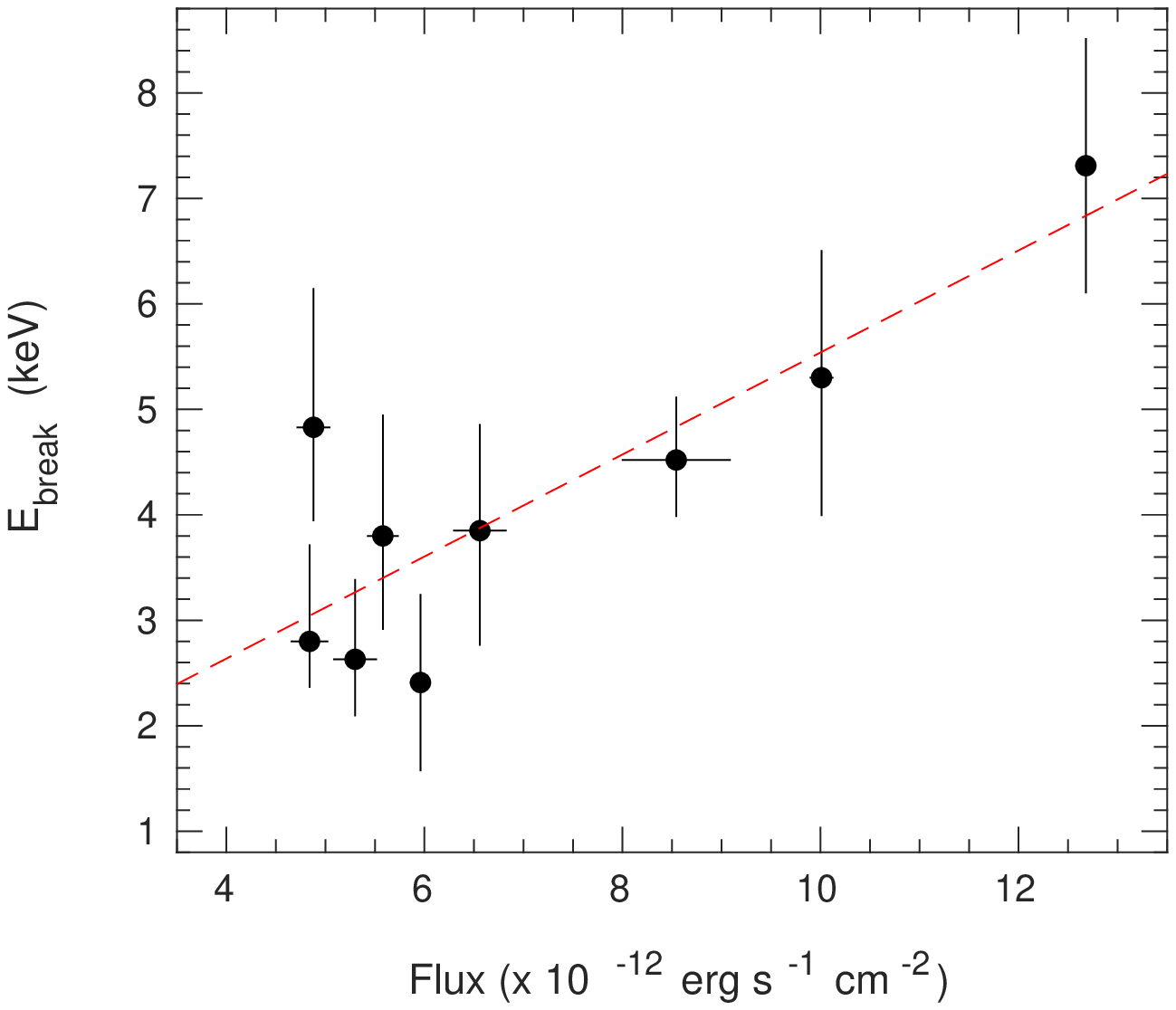}}\quad

\subfloat{\includegraphics[scale=0.4]{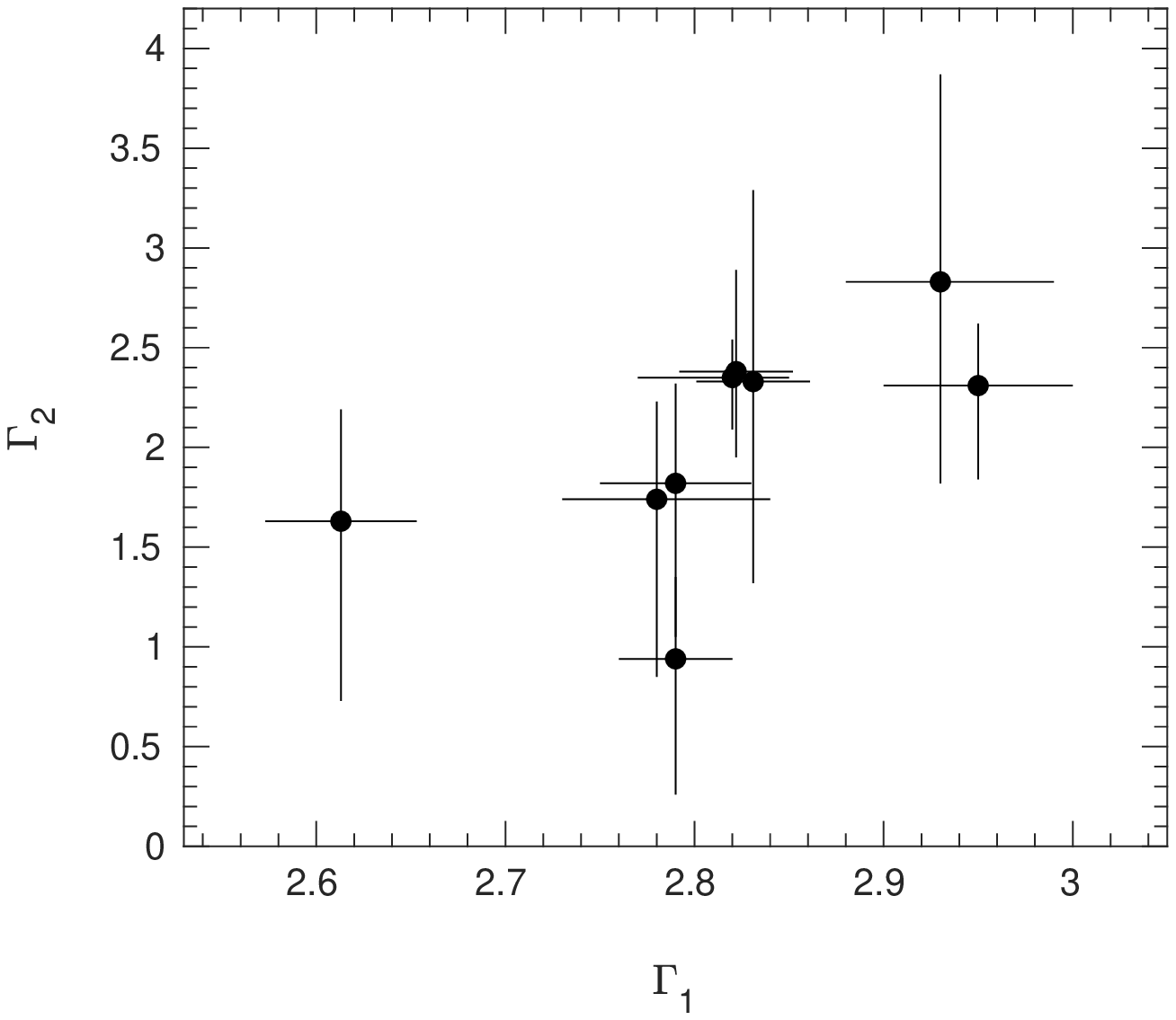} }}
\caption{Left: the ratio of the soft and hard X-ray spectral indices (SI) against the total flux derived from fitting a BPL to the average X-ray spectra. Middle: break energy against the the 0.6--10 keV total flux, with the black dashed line giving the least-square fit, showing a positive correlation between them. Right: variation of the hard SI against the soft SI. }
\end{figure*}

We repeated the spectral fitting for all the time intervals mentioned above (with BPL and LP models).
We first fitted the spectra with an absorbed BPL model in order to find out the break energy where the IC component becomes dominant over the synchrotron component and also to find the spectral slope of Compton-scattered spectra. Later, the E$_{break}$ values were used to set the upper limit for the synchrotron tail. It is important to note that the bandwidth of a spectrum plays a major role in constraining the curvature parameter precisely \citep{2004A&A...413..489M}. When we used only the X-ray spectra, we got a negative value of $\beta$ for intervals T31 $\&$ T32, while the parameters could not be well constrained for the interval T12. These points were immediately resolved when we included the optical/UV data points to the spectral fits. Thus, to achieve higher accuracy of this analysis, we combined the X-ray spectra with the averaged optical and UV data of the corresponding observations to each of the time intervals. We then fitted the synchrotron spectra from optical up to $E_{b}$ with an LP model to investigate the time-sensitive spectral behavior of the emission component. 
 
Due to the poor statistics at HEs, the photon index of the IC spectrum could not be well constrained while fitted with a BPL model. The same behavior was observed during the historical optical outburst of the source in 1998, reported by \citet{2000A&A...354..431T}. The results of these time-resolved spectral fits are presented in Tables 6 and 7. 

In Figure 9, we plot the spectral indices of both components and break energies as a function of total X-ray flux (left and middle panels). A rapid change of the break energy with respect to the total flux is evident from the middle panel. The red dashed line represents a linear fit to the data. The relation between the two spectral components is shown in the right panel of the figure, which indicates a positive correlation between the two emission components. Figure 10 shows the relationships between different spectral parameters and the synchrotron flux derived from the time-resolved spectral fit of the synchrotron component only with an LP model. Neither the spectral index ($\alpha$) nor the spectral curvature ($\beta$) appears to be related to the synchrotron flux (middle and right panels of the figure). However, there is a positive correlation between $\alpha$ and $\beta$ (left panel), as it should be for spectra having convex shapes. Variation of the curvature parameter and the flux of the curved synchrotron spectra with respect to the peak frequency are depicted in Figure 11 and 12, respectively.

\section{Discussion and Conclusions}

In this paper, we presented a detailed spectral and temporal analysis of the TeV-emitting intermediate-type blazar ON 231 (W Comae) with observations taken by the {\it XMM--Newton} and {\it Swift} satellites. Our study was mainly focused on the major X-ray outburst detected during the TeV $\gamma$-ray outburst in 2008 June. In order to understand the underlying physical processes, we carried out an X-ray spectral study in the energy range $0.6-10$ keV, along with an SED and correlation analysis using simultaneous optical and UV data.

The averaged X-ray spectra of the ISP blazar during the flaring event can be well represented by a BPL model as explained in Section 3. The presence of an upturn in all the spectra indicates that the crossing point where the IC emission becomes dominant over the synchrotron emission is located within the energy range of {\it Swift/XMM--Newton}. Thus, it allows us to investigate simultaneously both the HE tail of the synchrotron emission component and the low-energy side of the IC component. We find that the crossing point, i.e., the break energy, increases rapidly with the source flux, varying between $2.38$ keV and $5.33$ keV during the observing period (details are listed in Table 3). We find that the IC spectra were flatter than the synchrotron spectra and the total X-ray flux was significantly dominated by the synchrotron emission. Similar trends were found for other intermediate BL Lacs in previous studies \citep[e.g.,][]{1999A&A...351...59G, 2000A&A...354..431T,2003A&A...400..477T}. 


\begin{table*}
\centering
\caption{Result of time-resolved spectral analysis of the synchrotron component, from optical$-E_{b}$ with a log-parabolic power-law model}
\small
\begin{tabular}{lcccccccc} \hline \hline
ObsID&Int. &$\alpha^{a}$&$\beta^{b}$ & N$^{c}$ &$\chi_{r}^{2}$(dof)& $F_{sync1}^{d}$&  $F_{sync2}^{e}$& $\nu_{p}^{f}$ \\
 & & & &($\times$ 10$^{-03}$)& &  (erg cm$^{-2}$ s$^{-1}$) & (erg cm$^{-2}$ s$^{-1}$) & ($Hz$)\\\hline

0502211301&T11& 2.71$\pm$0.03& 0.14$\pm$0.01&  1.68$\pm$0.03&  1.22 (240)& 3.75$\pm$0.07$\times 10^{-12}$ &4.30$\pm$0.08$\times 10^{-12}$ & 7.34$\times 10^{14}$ \\ 
          &T12& 2.66$\pm$0.02& 0.16$\pm$0.01&  2.85$\pm$0.04&  0.95 (272)& 6.43$\pm$0.08$\times 10^{-12}$ &7.43$\pm$0.02$\times 10^{-12}$  & 2.03$\times 10^{15}$ \\ 
          &T13& 2.69$\pm$0.02& 0.18$\pm$0.01&  3.29$\pm$0.05&  1.07 (246)& 7.35$\pm$0.10$\times 10^{-12}$ &8.41$\pm$0.15$\times 10^{-12}$  & 3.22$\times 10^{15}$ \\ 
          &T14& 2.68$\pm$0.02& 0.19$\pm$0.01&  3.89$\pm$0.05&  1.22 (278)& 8.67$\pm$0.11$\times 10^{-12}$ &9.92$\pm$0.13$\times 10^{-12}$  & 4.27$\times 10^{15}$\\  
          &T15& 2.64$\pm$0.02& 0.19$\pm$0.01&  4.85$\pm$0.06&  1.21 (306)&10.92$\pm$0.14$\times 10^{-12}$ &12.59$\pm$0.16$\times 10^{-12}$ &5.42$\times 10^{15}$ \\  

0502211401&T21& 2.73$\pm$0.04& 0.19$\pm$0.02&  2.39$\pm$0.05&  1.31 (179)& 5.28$\pm$0.10$\times 10^{-12}$ &5.99$\pm$0.15$\times 10^{-12}$  & 2.77$\times 10^{15}$ \\ 
          &T22& 2.77$\pm$0.04& 0.18$\pm$0.02&  1.85$\pm$0.04&  1.17 (192)& 4.05$\pm$0.08$\times 10^{-12}$ &4.57$\pm$0.12$\times 10^{-12}$  & 1.96$\times 10^{15}$\\  
          &T23& 2.69$\pm$0.04& 0.15$\pm$0.02&  1.94$\pm$0.04&  1.13 (204)& 4.33$\pm$0.09$\times 10^{-12}$ &4.99$\pm$0.15$\times 10^{-12}$  & 1.35$\times 10^{15}$ \\

0502211201&T31& 2.63$\pm$0.03& 0.12$\pm$0.01&  2.33$\pm$0.04&  1.44 (252)& 5.33$\pm$0.10$\times 10^{-12}$ &6.24$\pm$0.12$\times 10^{-12}$  & 5.73$\times 10^{14}$ \\
          &T32& 2.59$\pm$0.04& 0.09$\pm$0.02&  1.93$\pm$0.04&  1.27 (177)& 4.49$\pm$0.11$\times 10^{-12}$ &5.35$\pm$0.17$\times 10^{-12}$  & 9.83$\times 10^{13}$ \\\hline

\end{tabular}     \\
Notes.
 $^{a}$ Photon index of the LP evaluated at 1 keV,
 $^{b}$ $\beta$ is the curvature parameter of the model, $^{c}$ Model normalization, \\
 $^{d}$ $0.6-4$ keV synchrotron flux,
 $^{e}$ Extended $0.6-10$ keV synchrotron flux,
 $^{f}$ SED peak frequency in Hz. 
\end{table*}

\begin{figure*} 
\centering
\mbox{\subfloat{\includegraphics[scale=0.39]{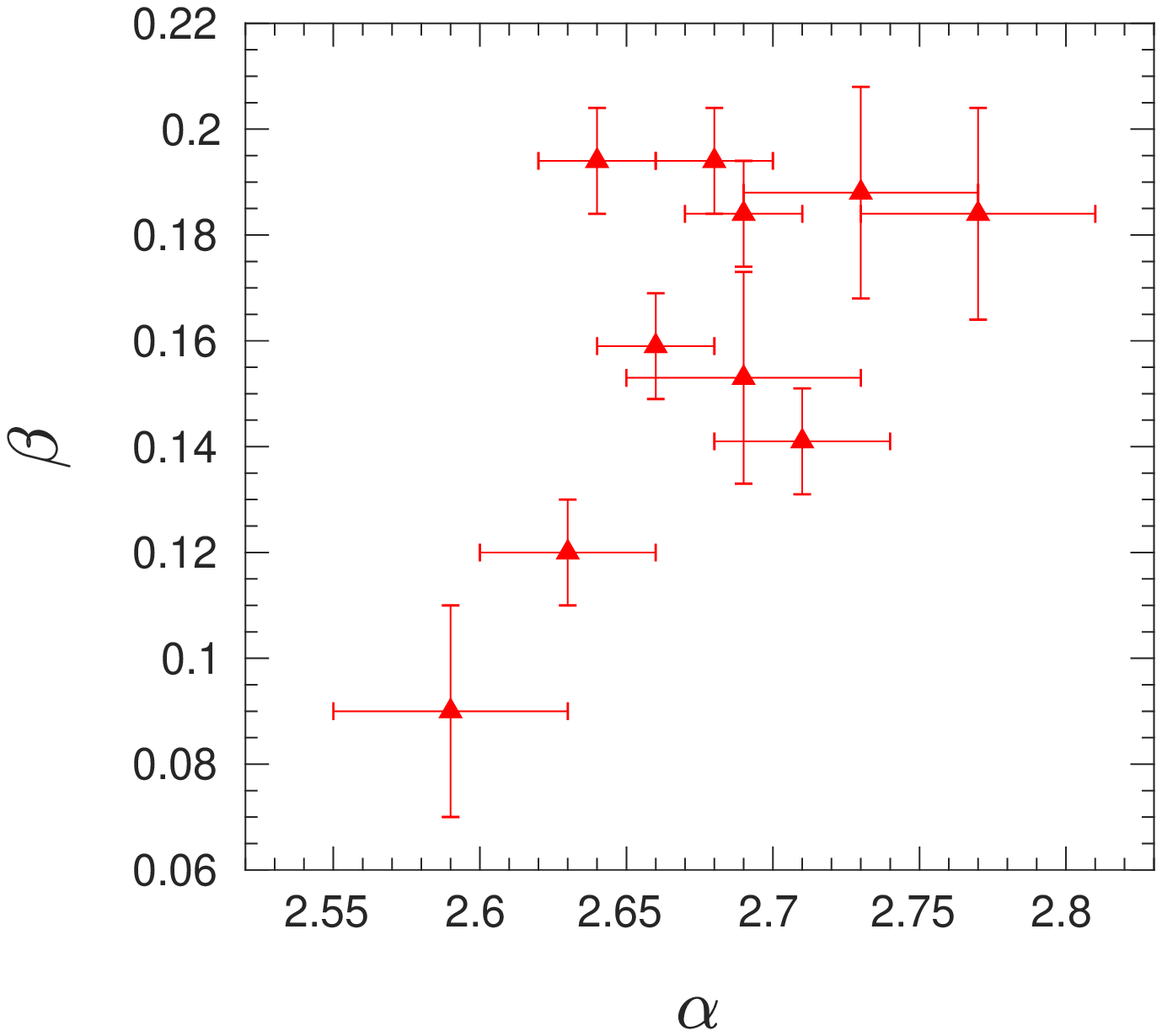}}\quad
\vspace*{.2cm}
\subfloat{\includegraphics[scale=0.39]{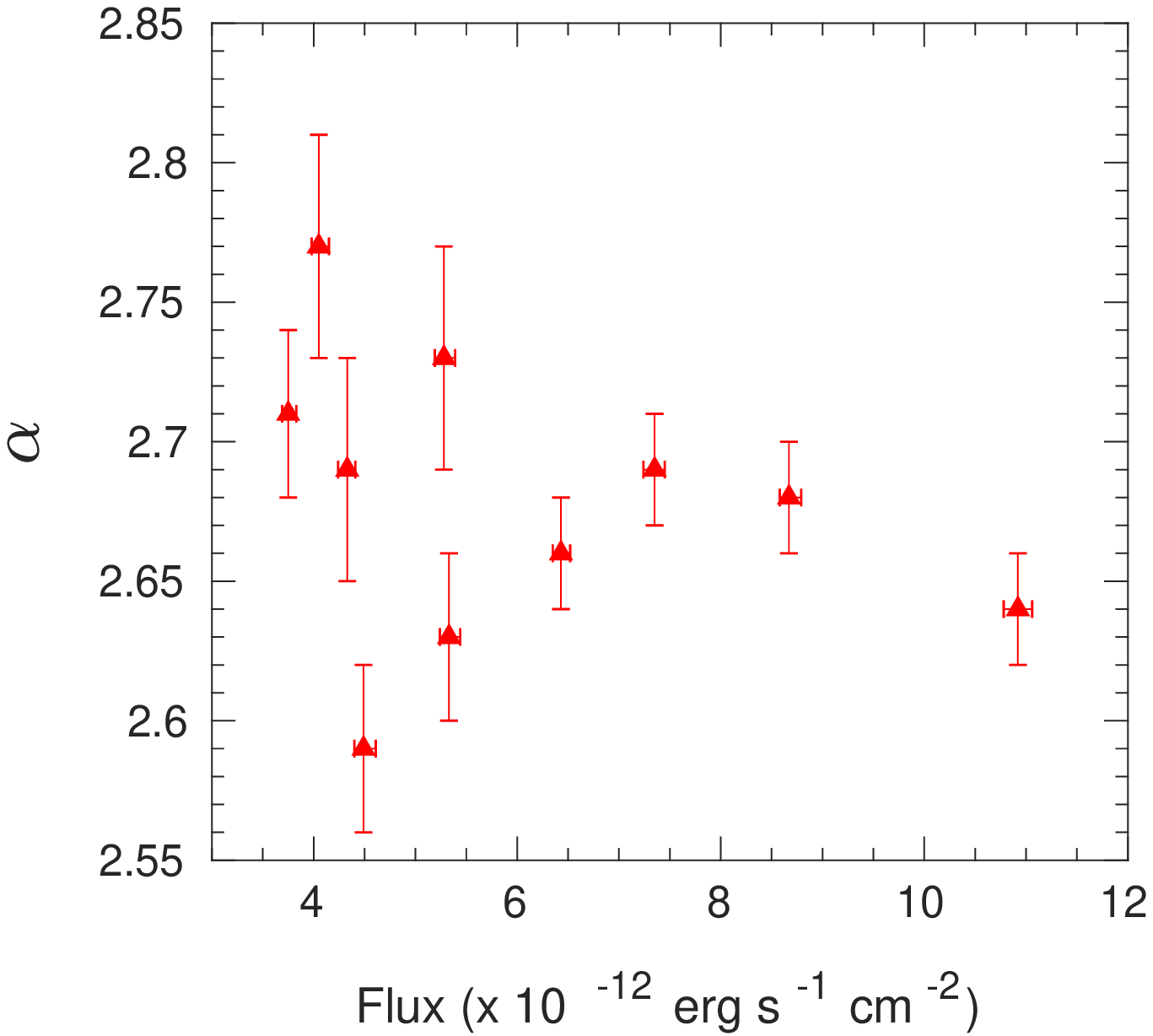}}\quad
\vspace*{.2cm}
\subfloat{\includegraphics[scale=0.39]{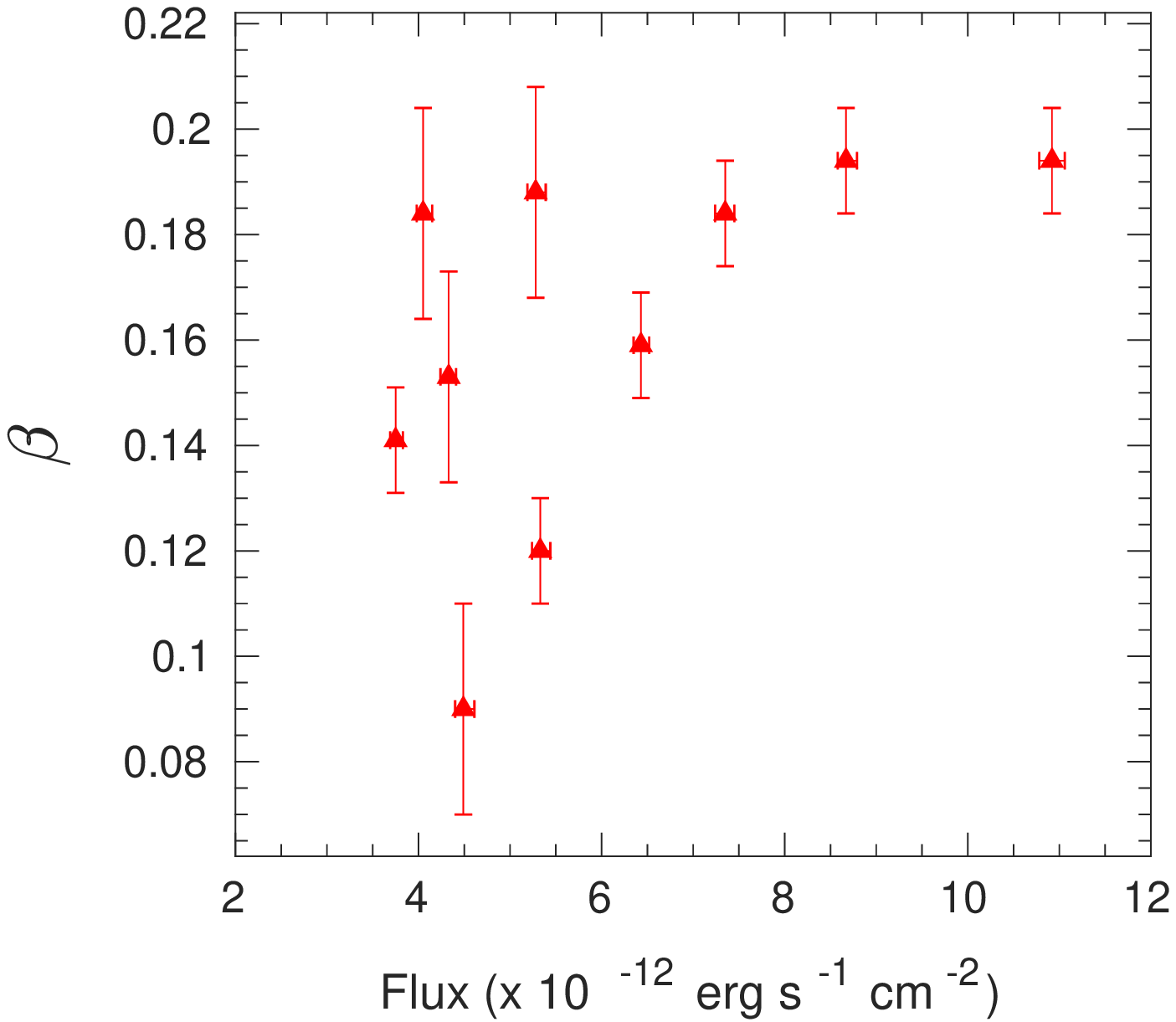} }}
\caption{Left: positive pattern between the curvature parameters and the spectral indices estimated from the time-resolved spectral fitting with an LP model. Middle and right: variation of the curvature parameters, $\beta$, and the spectral indices, $\alpha$, with the 0.6--4 keV synchrotron flux, respectively. No clear pattern is observed in the last two panels.}
\end{figure*}

Studies have found that excess of absorption of the soft X-ray photons might result in a curved spectrum that is not intrinsic in nature. Apart from the neutral hydrogen column density in our Galaxy, there could be local gas within the blazar itself that might cause an excess of absorption of the soft X-rays. The possible existence of intrinsic gas within TeV blazars surfaced with the detection of molecular carbon monoxide (CO) in millimetric surveys \citep{1993ApJ...403..663L, 2012MNRAS.424.2276F}. The additional absorption due to the presence of a significant amount of gas could lead to misinterpretation of the X-ray spectrum as intrinsically curved \citep{2013ApJ...770..109F}. Those authors investigated the topic for three well-known TeV blazars, including W Comae, using $0.3-10$ keV X-ray spectra and found no evidence for intrinsic gas in the source. Based on these studies, we believe that the spectra studied in this work are not affected by excess absorption and the curvatures we detected are intrinsic to the emission region.

In order to investigate the flux and spectral properties around the synchrotron peak, we have generated SEDs of the source using simultaneous optical, UV, and X-ray data. Study of blazar SEDs is useful to understand the emission mechanisms and is an excellent diagnostic tool to check the relevance of different existing jet models. Each section of the multiwavelength SED can provide significant information about the corresponding emitting regions. In our analysis, we have found that the broadband spectra of the flare follow a curved shape which can be well represented by an LP function (discussed in section 4). This function has only three variables and has been extensively used to fit SEDs of high-energy BL Lac objects owing to its simplicity \citep{1962SvA.....6..317K, 1986ApJ...308...78L, 2004A&A...413..489M, 2006A&A...448..861M, 2011ApJ...739...66T, 2009A&A...501..879T}. A simple explanation for the origin of log-parabolic energy spectra can be given in terms of a statistical acceleration process for relativistic electrons where the probability of acceleration is a decreasing function of electron energy \citep{2004A&A...413..489M}. However, other possible interpretations are present in the literature \citep[e.g.,][]{2007A&A...466..521T}. Generally, curved blazar spectra reflect radiative aging of the emitting particles that can be assumed to have a single PL distribution during their injection. As shown in Section 4 (see Fig.\ 5a), we find a significant relation between the spectral curvature ($\beta$) and slope ($\alpha$). This is an indication that the LP curved spectra are intrinsically related to particle acceleration processes, and this trend can also be discerned in the time-resolved spectra (Fig.\ 10, left). It also appears that as the SED peak moves to higher energies, the spectra become more curved (top right panel of Fig.\ 5).

In this burst event, the source showed very high flux variation in the X-ray band, with variability amplitudes ranging between $27\%$ and $38 \%$ (Table 1), while the optical/UV emissions showed milder variation, with $F_{var}$ $\approx$ $\sim 4-8 \%$ (Table 2). The UV flux of the source changes between values of $\sim 2.8 \times 10^{-11}$ erg s$^{-1}$ cm$^{-2}$ (taken from {\it Swift}-UVOT; \cite{2009ApJ...707..612A} and $\sim 2.1 \times 10^{-11}$ erg s$^{-1}$ cm$^{-2}$ (recorded by {\it XMM--Newton--OM}), respectively. The relationship between X-ray variability amplitudes and their corresponding energies is somewhat more complex than the usually observed trends. We see three different patterns for the three {\it XMM} observations in Fig.\ 6. A linearly opposed trend of the energy-dependent variability is observed at the beginning, which takes on a curved shape toward the end of the event. Another important point is that the source emitted substantially more X-rays in soft bands, below $3-4$ keV (Figure 2), while above that energy the photon flux decreased to $<1$ count s$^{-1}$. 

In our DCF analysis, we found that all the X-ray sub-bands below the break point, which was generally around $4$ keV, are correlated to each other without any significant time delay. This indicates that photons in these energy bands were emitted via the same physical mechanisms that took place in a single region. The emission in the range $0.3-0.5$ keV that lagged behind the higher-energy emission at $4-10$ keV by $\sim 910$ s implies that the distance between these two emitting regions must be small. This type of lag can be interpreted through synchrotron cooling of the high-energy electrons from different parts of an inhomogeneous relativistic jet: the higher frequencies could be emitted from the inner and denser parts of the jets, while the lower frequencies could be emitted from comparatively more distant and less dense parts or from greater distances from a shock front \citep{1996A&AS..120C.537M}. This also may be a result of material propagating through a rotating helical path \citep{1999A&A...347...30V}. However, the degree of correlation is not very strong, and the lag measurement errors are not negligible, so this observed weak relation could also be due to the remnants of synchrotron emission still present in the IC component ($4-10$ keV).

We are viewing blazars nearly directly down the jet, so the observed variability is generally believed to originate inside the jet, which is significantly amplified by relativistic beaming. The multiwavelength flux variations in blazars are often interpreted in terms of shocks traveling down the relativistic jets pointing close to our line of sight \citep[e.g.,][]{1985ApJ...298..114M}. Changes of the jet angle, i.e., varying jet direction with our line of sight \citep{2017Natur.552..374R} or other geometrical effects occurring within the nonthermal jet \citep[e.g.,][]{1992A&A...255...59C, 1992A&A...259..109G} provide other possible explanations. Apart from that, accretion-disk-induced fluctuations during advection of matter into the jet could play an important role in blazar variability by changing the physical parameters of the jet, e.g., magnetic field, the velocity of the emitting gas, or its density. The important point is that, thanks to extreme Doppler boosting, even a tiny variation in these parameters may cause a huge difference in the observed emission and the variability timescale \citep[e.g.,][]{1978ApJ...221L..29B, 2003ApJ...586L..25G}. The above discussed scenarios have been successfully used to interpret the synchrotron emission, which extends up to UV ranges and sometimes up to X-ray regimes. The main hard X-ray emission mechanism in blazars is not the synchrotron process but the IC scattering of low-energy photons off highly energetic electrons in a hot plasma. Different models have been used to describe the observed high-energy emissions in blazars. One possible scenario is that the soft photons could be Comptonized by an accretion disk corona residing above the disk or by the relativistic electrons in the jet (EC or SSC models). However, the whole picture of X-ray emission in most blazars is still somewhat fuzzy and demands further investigations. 

It is obvious from the SED analysis that the optical/UV to soft X-ray emissions were generated by a synchrotron radiation process within the jet, and as reported in Section 5.2, we found soft lags between the UV and low-energy X-ray bands (Table 5). \citet{2002MNRAS.337..609Z} showed that the time lag observed between different synchrotron bands can provide a way to estimate the magnetic field strength of the emitting regions as follows:
\begin{equation}
B\delta^{1/3} =209.91 \times \left (\frac{1+z}{E_{\rm
l}}\right)^{1/3}
    \left [\frac{1 - (E_{\rm l}/E_{\rm h})^{1/2}}
        {\tau_{\rm soft}} \right ]^{2/3} \quad {\rm G}\,,
\label{eq:soft}
\end{equation}
where $\tau_{\rm soft}$ is the observed soft lag (in seconds) between
the low-energy ($E_{\rm l}$) and high-energy ($E_{\rm h}$) bands (in keV), $z$
is source's redshift, $B$ is magnetic field strength in Gauss, and $\delta$
is the bulk Doppler factor of the emitting region. If we use the maximum time lag from DCF analysis, which is $\tau_{\rm soft} \sim -3250$~s between the UV and $0.3-0.5$ keV bands, the magnetic field estimated from the above equation returns a value $B\delta^{1/3} \sim 5.67$~G. A focused SED modeling of the 2008 June TeV outburst event with contemporaneous multiwavelength observations was presented by \citet{2009ApJ...707..612A}, where they used a quasi-equilibrium leptonic one-zone jet model \citep{2002ApJ...581..127B}. They modeled the broadband emission using a pure SSC model and a combined SSC $+$ EC model, assuming a Doppler factor of $20$, and measured the magnetic field strength having values of $0.25$ G and $0.35$ G from the respective models. They found that an SSC+EC model gives a more satisfactory explanation of the SED and reported that the external radiation field producing the EC emission at VHE $\gamma$-rays is most likely the IR emission originating from the dusty torus.

In the case of TeV-emitting blazars, a harder-when-brighter trend of the X-ray spectrum has been reportedly observed from time to time \citep[see][and references therein]{2017ApJ...841..123P, 2018MNRAS.480..879M}. However, we do not see any clear trend between the X-ray flux and the spectral indices in our time-resolved spectral study. A shifting of the synchrotron peak frequency by around two orders of magnitude was observed throughout the event.  

In contrast with other TeV blazars, we have found a significant positive correlation between the peak frequency ($\nu_{p}$) and spectral curvature ($\beta$) from both the overall SED (Fig.\ 5) and the time-resolved spectral analysis, shown in Figure 11. In general, the curvature parameters and the peak energies are anticorrelated with each other \citep[see][and references therein] {2004A&A...413..489M, 2007A&A...466..521T, 2009A&A...501..879T, 2006A&A...453...47K}, which have been explained in the framework of energy-dependent acceleration mechanisms where the acceleration probability decreases with the particle energy \citep{2004A&A...413..489M} or through a stochastic acceleration process of relativistic particles undergoing momentum diffusion \citep{1962SvA.....6..317K, 2007A&A...466..521T}.

 \begin{figure}
\centering
\includegraphics[scale=0.48]{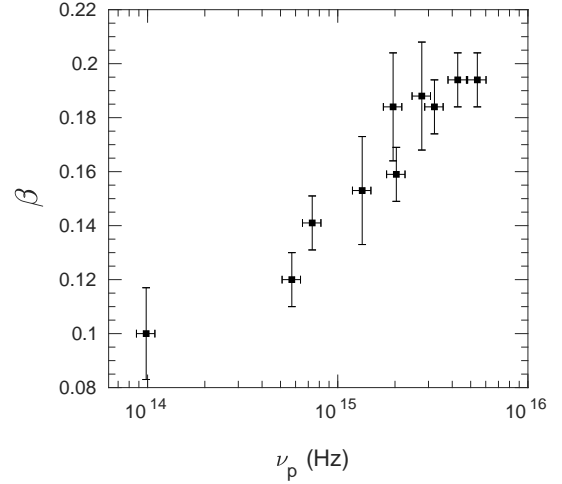}
\caption{Synchrotron spectra become more curved with an increasing peak frequency of the SED. Data points are extracted from time-resolved spectral analysis of the synchrotron component.}
\end{figure}   

\begin{figure}     
\centering
\includegraphics[scale=0.48]{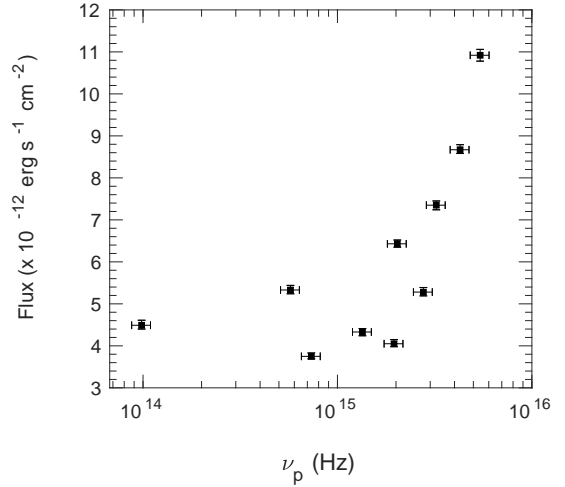}
\caption{Synchrotron peak frequency vs. $0.6-4$ keV flux gives a weak positive correlation.}
\end{figure}

The $\nu_{p}-\beta$ relation in blazars plays an important role in understanding the physical mechanisms and parameters governing the evolution of the particle distribution. The $\nu_{p}$--$\beta$ trend we observed in Fig.\ 11 can be explained in the framework of stochastic acceleration and cooling processes \citep{2011ApJ...739...66T}, where the magnetic field plays an important role in the evolution of the spectral parameters. These authors investigated the evolution of the curvature parameter of the electron distribution resulting from momentum diffusion in a stochastic acceleration process. Under the SSC scenario, and assuming that the electrons are injected into the acceleration region with a quasi mono-energetic spectrum, they showed that as long as the magnetic field is weak, the evolution of the particles around the peak is dominated by the acceleration process, while it is driven by cooling for higher values of $B$. In this context, the $\nu_{p}-\beta$ trend observed in Fig.\ 11 shows the role played by {\it B}-field in the evolution of the particle distribution. In the case of $\delta$-function approximation of the synchrotron emission \citep{1986rpa..book.....R}, \citet{2011ApJ...739...66T} showed that for $B \leq 0.2$ G, the spectral curvature is almost unaffected by the magnetic field and shows a stable value of $\beta$ with $\nu_{p}$, but for $B \geq 0.2$ G, the curvature parameter increases as a function of $\nu_{p}$. In our study, we find the minimum magnetic field strength near the emitting region inside the jet to be $\sim 5.67\delta^{-1/3}$ G from Eqn.\ (8). If we consider $\delta = 20$ as \citet{, 2009ApJ...707..612A} did in their study, we get a magnetic field intensity $\approx 2$ G, which is higher than the above threshold value. Thus, the emission during the event was most likely governed, or at least strongly affected by, the magnetic field. Another important point that can be drawn from Fig.\ 11 relates to the stage of the particle distribution, as the pattern in the figure is observed when the emission processes have closely approached the equilibrium energy, where the acceleration process is fully compensated by the cooling. From our analysis, we conclude that the X-ray flaring event observed during the TeV outburst likely was driven by the {\it B}-field.

The main findings of this work are summarized as follows 
\begin{itemize}

\item {\it Swift}--XRT and {\it XMM--Newton} EPIC-pn and OM observations have been used to study the X-ray outburst phase of the ISP blazar ON 231 observed in 2008 June.

\item The average X-ray spectra show the superposition of both the synchrotron and IC components, which can be well explained by a BPL model. The crossing point, where the two components contribute equally, varies rapidly from $2.38$ to $5.33$ keV with flux.

\item Although a BPL model can explain the overall X-ray spectra, an LP function was required to precisely describe the synchrotron part of the broadband SEDs. The spectral curvature and slope are tightly related to each other, as it should be for a curved electron distribution. During the event, the SED peak shifted by two orders of magnitude, from $\sim 10^{14}$ Hz to $10^{16}$ Hz (see Table 7 and Fig.\ 11).

\item We find significant flux variation on intraday timescales, with $F_{var}$ ranging from $27\%$ to $38\%$ in the X-ray band. The observed variability was mainly due to the variation of the synchrotron component. However, the simultaneous optical/UV fluxes only showed mild variations.

\item All the soft X-ray bands, which are produced by synchrotron emission, were correlated to each other with zero lag, while a soft lag of $-910$ s was detected between the $0.3-0.5$ keV and $4-10$ keV bands. A maximum delay of $1$ hr was detected between the UV and soft X-ray bands, which gives a {\it B}-field strength of $\sim 5.67 ~\delta^{-1/3}$ G in the jet.

\item Time-resolved spectral analysis of multiepoch data revealed that the break energy of X-ray spectra actually varies significantly from $2.4$ to $7.3$ keV and the total flux changes between $4.9 \times 10^{-12}$ erg cm$^{-2}$ s$^{-1}$ and $12.7 \times 10^{-12}$ erg cm$^{-2}$ s$^{-1}$. 

\item Finally, the X-ray outburst of ON $231$ in $2008$ June appears to have been a magnetic-field-driven flaring event where the evolution of the particle distribution was closely approaching the equilibrium energy via stochastic acceleration and cooling processes.
\end{itemize}

\begin{figure*}[b]
  \centering
  \subfloat[][]{\includegraphics[scale=0.5]{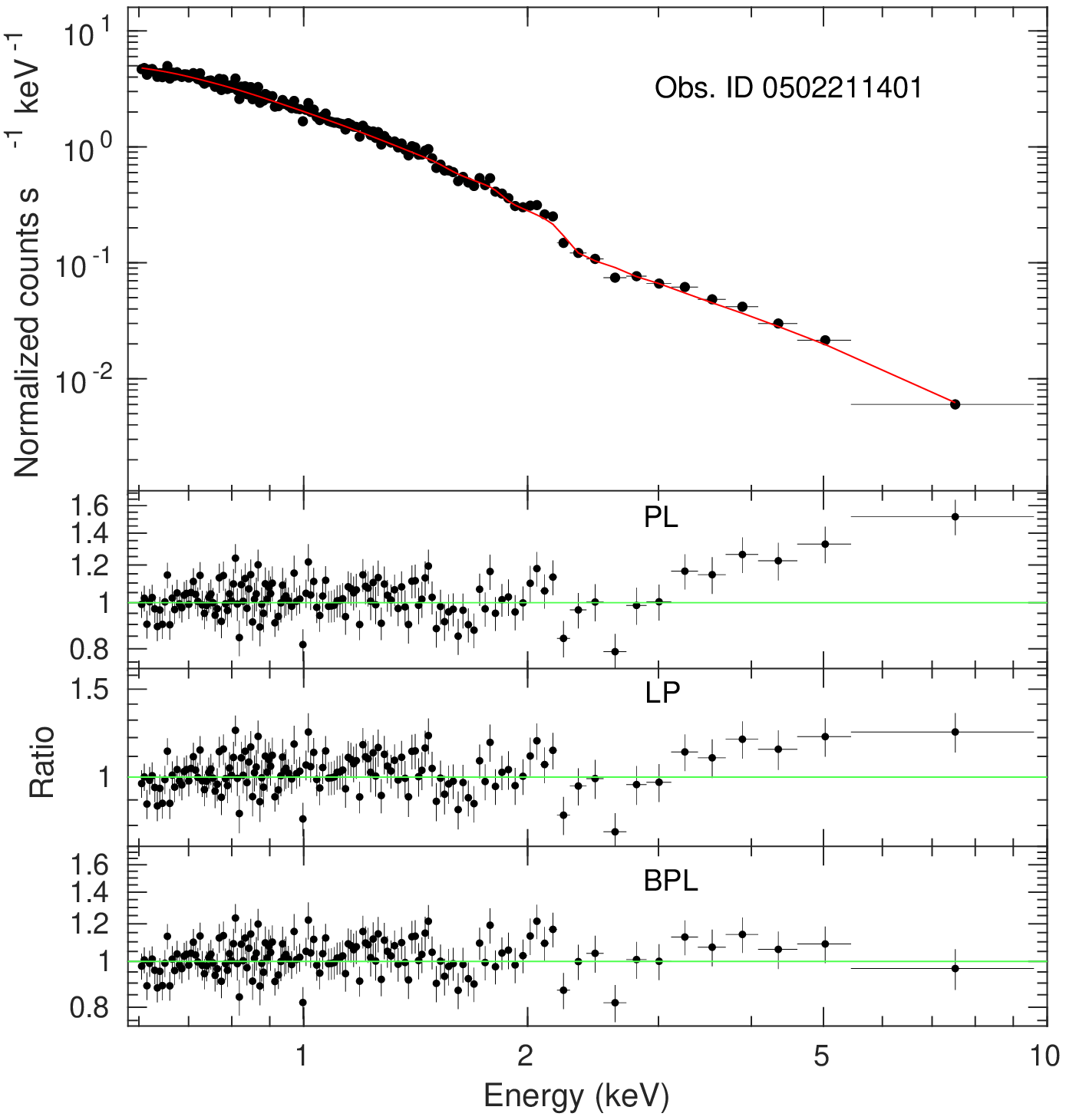}}\quad
  \subfloat[][]{\includegraphics[scale=0.5]{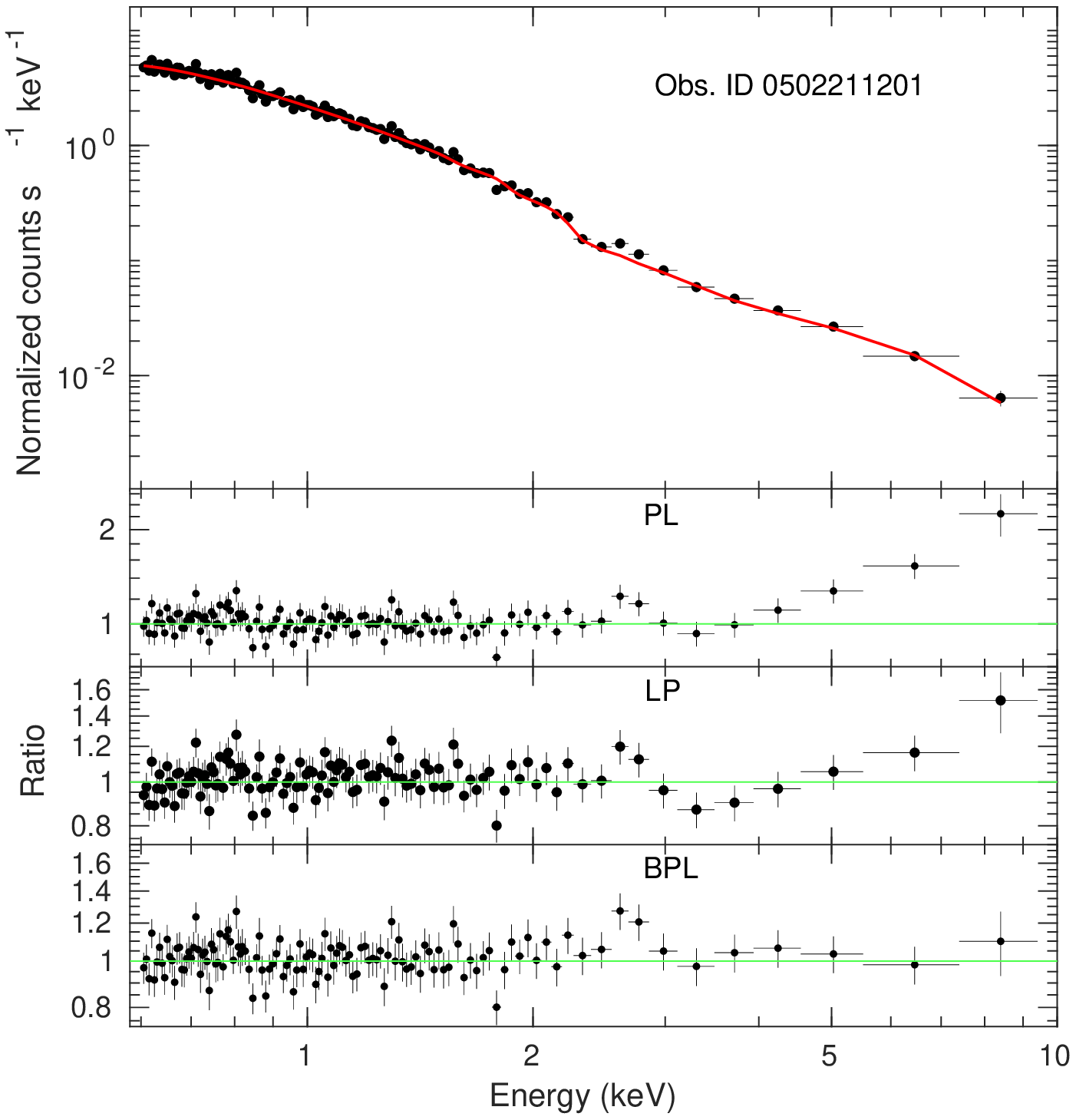}}\\
  \subfloat[][]{\includegraphics[scale=0.5]{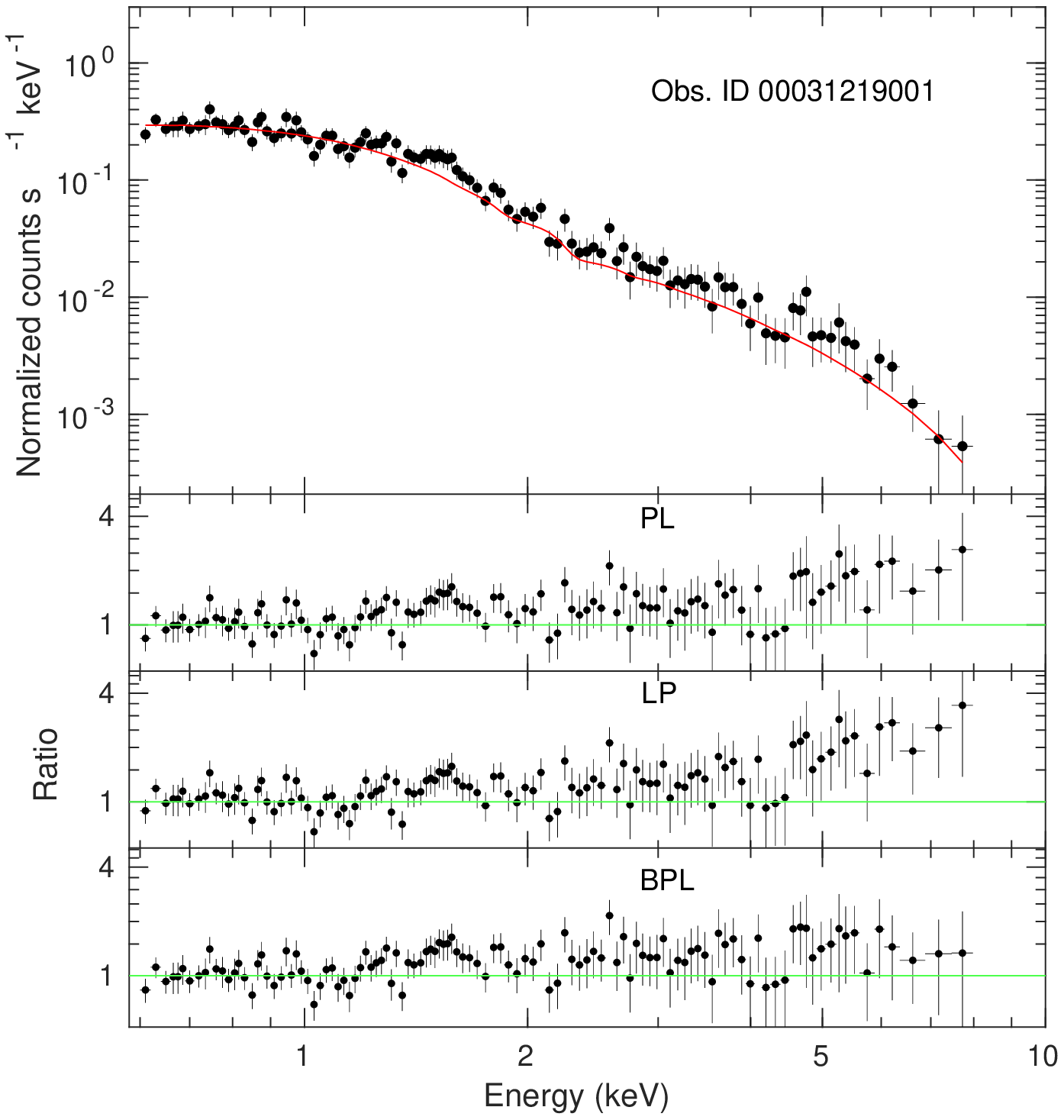}}\quad
  \subfloat[][]{\includegraphics[scale=0.5]{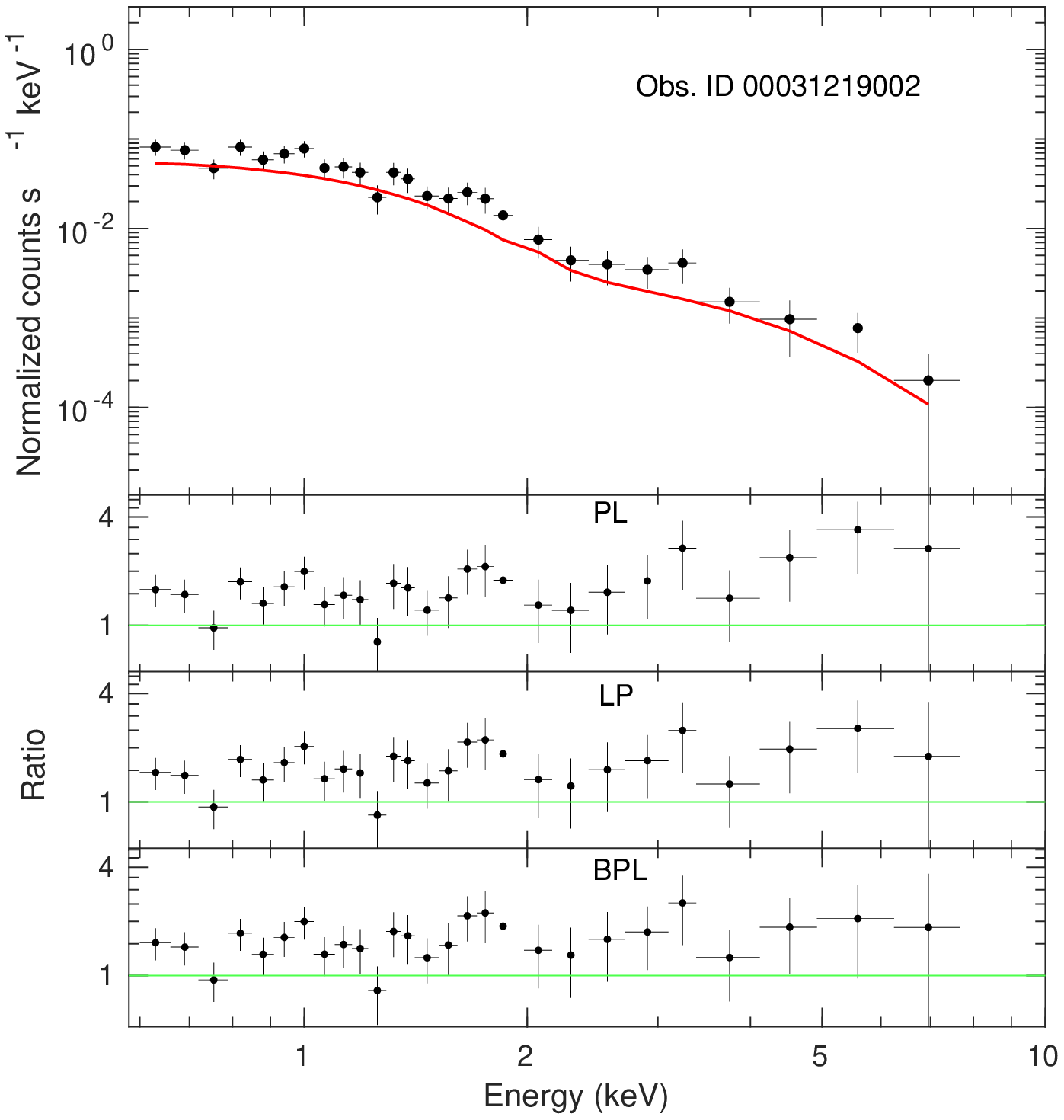}}\\
  
\caption{The $0.6-10$ keV average X-ray spectra as observed by the {\it XMM-Newton} \& {\it Swift} satellites are displayed with data-to-model ratios for three different spectral models. A positive residual in the HE tail is observed while the spectra are fitted with PL or LP models. The best fit is obtained with an absorbed BPL model.}
\end{figure*}

\acknowledgments 

We thank the anonymous referee for constructive comments and suggestions. This research has made use of data, software, and web tools obtained from NASA's High Energy Astrophysics Science Archive Research Center (HEASARC), a service of the Goddard Space Flight Center and the Smithsonian Astrophysical Observatory. This research has made use of {\it XMM--Newton} data, an ESA science mission and the XRT Data Analysis Software (XRTDAS) developed under the responsibility of the ASDC, Italy.

\software{SAS \citep[v17.0.0;][]{2004ASPC..314..759G}, HEAsoft (HEASARC 2014), XSPEC \citep[v12.8.2;][]{1996ASPC..101...17A} } 

\bibliographystyle{aasjournal}

\bibliography{myref}

\end{document}